%% file: mb1069.tex
\documentstyle[psfig]{mn}
\footnotesize
\newdimen\digitwidth    %define ! a one digit width for tables
\setbox0=\hbox{\rm0}
\digitwidth=\wd0
\catcode`!=\active
\def!{\kern\digitwidth}
\normalsize

\title[Parkes Multibeam Pulsar Survey] {The Parkes Multibeam Pulsar
Survey --  II.  Discovery and Timing of 120 Pulsars}
\author[D. J. Morris et al.]
{D. J. Morris,$^1$
G. Hobbs,$^1$
A. G. Lyne,$^1$
I. H. Stairs,$^{2,1}$ 
F. Camilo,$^3$
\newauthor
R. N. Manchester,$^4$
A. Possenti,$^5$
J. F. Bell,$^4$ 
V. M. Kaspi,$^{6,7}$
N. D'Amico,$^{5,8}$ 
\newauthor
N. P. F. McKay,$^1$
F. Crawford$^{6,9}$
and
M. Kramer$^1$
\\
$^1$ University of Manchester, Jodrell Bank Observatory, Macclesfield,
Cheshire SK11~9DL \\
$^2$ National Radio Astronomy Observatory, PO~Box~2, Green Bank,
WV~24944, USA \\
$^3$ Columbia Astrophysics Laboratory, Columbia University, 550 West
120th Street, New York, NY~10027, USA \\
$^4$ Australia Telescope National Facility, CSIRO, PO~Box~76, Epping
NSW~1710, Australia \\
$^5$ Osservatorio Astronomico di Bologna, via Ranzani 1, 40127~Bologna,
Italy \\
$^6$ Massachusetts Institute of Technology, Center for Space Research,
70 Vassar Street, Cambridge, MA~02139, USA \\
$^7$ Physics Department, McGill University, 3600 University Street,
Montreal, Quebec, H3A~2T8, Canada \\
$^8$ Cagliari Astronomical Observatory, Loc. Poggio dei Pini, Strada
54, 09012 Capoterra (Ca), Italy \\
$^9$ Department of Physics, Haverford College, Haverford, PA 19041,
USA \\
}

\date{2002 April 05}
\begin{document}

\maketitle
\newcommand{\setthebls}{
%                 de-comment this line for double spacing:
%\baselineskip=20pt
}

\setthebls

\begin{abstract} 
The Parkes multibeam pulsar survey is a sensitive survey of a strip of
the Galactic plane with $|b|<5\degr$ and $260\degr < l < 50\degr$ at
1374\,MHz. Here we report the discovery of 120 new pulsars and
subsequent timing observations, primarily using the 76-m Lovell radio
telescope at Jodrell Bank.  The main features of the sample of 370 published
pulsars discovered during the multibeam survey are described.  Furthermore, 
we highlight two pulsars:  PSR~J1734$-$3333, a
young pulsar with the second highest surface magnetic field strength
among the known radio pulsars, $B_s = 5.4\times10^{13}$\,G, and
PSR~J1830$-$1135, the second slowest radio pulsar known, with a
6-s period.
\end{abstract}


\section{INTRODUCTION}\label{sec:intro}

The Parkes multibeam survey is an on-going survey of a $10\degr$-wide
strip along the Galactic plane ($|b|<5\degr$ and $l=260\degr$ to
$l=50\degr$). The survey aims to detect a large sample of pulsars for
population studies (more than 600 have been discovered so far) as well
as find individual objects of interest including binary pulsars, young
pulsars, EGRET source associations and high magnetic field pulsars. The
survey uses a 13-beam receiver on the 64-m Parkes radio telescope,
receiving two polarisations per beam over a 288-MHz bandwidth centred
on 1374\,MHz.  The survey's receiver system, data acquisition and
control procedures, offline data analysis and calculations of
sensitivity are published in Manchester et~al.~(2001) \nocite{mlc+01}
(hereafter Paper~I) along with  the discovery and timing parameters for
100 pulsars with declinations $<-35\degr$.  Timing observations of the
majority of new pulsars with declinations $>-35\degr$ are made
using the Jodrell Bank 76-m Lovell radio telescope.  Positional and
rotational parameters for 120 high-declination pulsars are given in
this paper along with corresponding derived parameters and averaged
pulse profiles\footnote{The data release policy and results for the
survey may be found at
http://www.atnf.csiro.au/research/pulsar/pmsurv.}.  The parameters for
a further 150 pulsars are given in Bell et al. (in preparation),
hereafter Paper~III.

In order to facilitate follow-up observations, it is desirable to
determine a better position estimate than that available from the survey
data.  For this reason, a technique known as `gridding' is carried out
on newly discovered pulsars before long-term timing begins.  This
technique, introduced in Paper~I, is described in detail here.  We also
describe the Jodrell Bank observing system, summarise some interesting
properties of the sample of 370 pulsars published here and in Papers~I
and III, and highlight two pulsars: PSR~J1734$-$3333, a young pulsar
with a high surface magnetic field and PSR~J1830$-$1135, a pulsar with
a 6-s period.

\section{TIMING OBSERVATIONS AND ANALYSIS}\label{sec:timing}

\subsection{Gridding}
\begin{figure*}
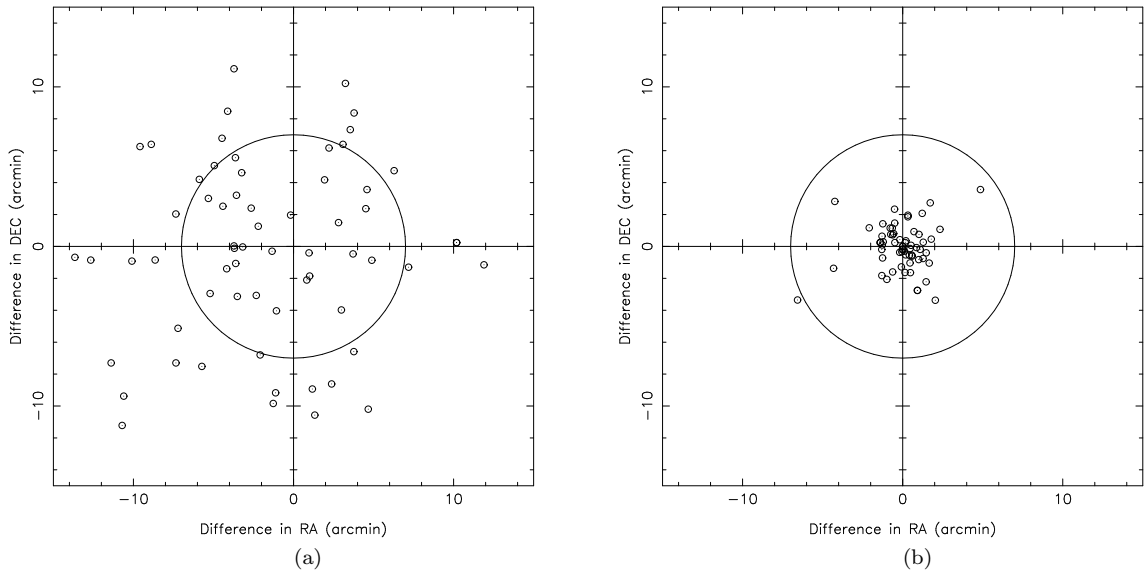
 
\begin{center}\begin{minipage}[t]{80mm}
\begin{center}\psfig{file=mb1069fig1a.ps,width=70mm,angle=-90} (a)\end{center}
\end{minipage}
\begin{minipage}[t]{80mm}
\begin{center}\psfig{file=mb1069fig1b.ps,width=70mm,angle=-90} (b)\end{center}
\end{minipage}\end{center}
\caption{In (a) each point represents the offset between the initial
candidate position and the precise position in right ascension (RA) and
declination (DEC) resulting from one year of timing observations for
the 82 pulsars published here and in Paper~I that have been gridded;
(b) shows the same for the positions given by the gridding
observations.  The circle represents the half-power beamwidth of the
Parkes telescope. }
\label{fg:plots}
\end{figure*}

The centre of the discovery beam defines the initial position given to
a pulsar candidate.  If the pulsar is offset from the centre of the
$14\farcm4$ beam then it will be detected with a lower signal-to-noise
ratio (S/N) than for an equivalent observation centred on the pulsar.
Furthermore, if the pulsar is to be observed with the narrower beam of
the Lovell telescope then the S/N may be further reduced, hence the
most accurate position estimate available should be used until the
precise position is determined from timing. The use of a more accurate
position maximises the S/N and therefore reduces the amount of
telescope time required to obtain a pulse arrival time.  Improved knowledge of
the position also reduces the density of observations required through
the year to obtain a coherent solution without period ambiguities in
the timing residuals.

In order to improve the position estimate,  a `gridding procedure' is
carried out using the Parkes telescope.  The original discovery
position is re-observed (with a shorter integration time of a few
minutes --- the exact time depending on the S/N in the original
discovery) along with four other integrations of similar durations at
positions $9'$ to the North, South, East and West.  The peak of a
Gaussian beam shape (with a half-power width of $14\farcm4$) fitted to
the S/Ns in these consecutive observations determines the pulsar
position with an uncertainty substantially smaller than a single
beamwidth.  Fig.~\ref{fg:plots}a contains a scatter plot of the offset
between the initial candidate position and the precise position
determined after one year of timing observations for all of the pulsars
with gridding observations (many of the earliest candidates were not
`gridded').  Many pulsars lie outside the half-power beamwidth:  the
root-mean-square (rms) positional uncertainty is $7\farcm2$. However,
the rms positional error of $1\farcm6$ for the differences between the
gridded and precise positions (Fig.~\ref{fg:plots}b) is well within
the beamwidth, and the typical area of uncertainty has been reduced by
more than an order of magnitude.

\subsection{Timing at Jodrell Bank}

For most of the timing observations, we use the Lovell telescope at
Jodrell Bank at a frequency near 1400\,MHz.  We employ a dual-channel
cryogenic receiver system sensitive to the two hands of circular
polarisation.  Each hand of polarisation is down-converted, fed through
a multichannel filterbank and digitized.  Until August 1999, the
filterbank consisted of $2\times32\times3$\,MHz channels centred on
1376\,MHz.  Subsequent observations use a filterbank with
$2\times64\times1$\,MHz channels centred on 1396\,MHz.

In the observing procedure, data are dedispersed and folded on-line
according to the pulsar's dispersion measure (DM) and topocentric
period.  The folded pulse profiles for each polarisation are stored for
subsequent analysis.  The duration of each observation is between 6 and
36 minutes (depending upon the flux of the pulsar) and is divided into
a number of sub-integrations, each lasting between one and three
minutes.  During offline data processing, the polarisations are
combined to produce the total intensity and the sub-integrations are
added to produce a final averaged profile for the observation.  Pulse
times-of-arrival (TOAs) are determined by cross-correlating the profile with a
template of high S/N.  These TOAs are corrected to the solar
system barycentre using the Jet Propulsion Laboratory DE200
solar-system ephemeris \cite{sta82}.  The positional and rotational
parameters for the pulsar are then obtained by model-fitting the
TOAs using {\sc tempo}\footnote{See http://pulsar.princeton.edu/tempo.}.

\input{mb1069tb1a.tex}
\addtocounter{table}{-1}
\input{mb1069tb1b.tex}
\addtocounter{table}{-1}
\input{mb1069tb1c.tex}

\input{mb1069tb2a.tex}

\addtocounter{table}{-1}
\input{mb1069tb2b.tex}
\addtocounter{table}{-1}
\input{mb1069tb2c.tex}

The search code produces an initial estimate of a pulsar's DM, and
observations at Jodrell Bank use a hardware dedisperser giving an
averaged pulse profile at a particular DM.  In order to make a more
precise measurement, we use either the discovery or the gridding
observations that used the 288-MHz bandwidth available at Parkes. The
band is divided into four sub-bands and the DM is obtained by fitting
to the TOAs of the pulses determined from each sub-band.

The flux density scale at Jodrell Bank is calculated by switching a
noise diode on and off between sub-integrations.  The noise diode is
calibrated using standard flux calibrators such as 3C295.  The pulsar
flux density for a single observation is then determined as the mean
flux density in the profile averaged over the pulse period.

Included in this paper are ten pulsars with low flux densities and
with declinations $>-35\degr$ that were timed at the Parkes telescope
because of the higher sensitivity afforded by the use of the larger
bandwidth.  The observing and data-reduction techniques used at Parkes
are described in Paper~I.  A few pulsars were observed with both
instruments, but for consistency, only Jodrell Bank data were used to
produce the pulse profiles.  Similarly, nine pulsars were timed on a
monthly basis with the 305-m Arecibo telescope, using a
$2\times128\times0.0625$\,MHz filterbank centred on 1400\,MHz.  These
pulsars were observed for 5-minute or shorter integrations, with the
polarisations summed and the data folded on-line using the known DM and
the predicted topocentric pulse period.  
TOAs were determined using a standard template as for the Jodrell Bank
observations.

\section{DISCOVERY AND TIMING OF 120 PULSARS}

The two tables giving positional, discovery and flux parameters
(Table~\ref{tb:posn}\nocite{lkm+00}) and rotational parameters
(Table~\ref{tb:per}) take the same form as the equivalent tables in
Paper~I.  In summary, Table~\ref{tb:posn} lists the positional
properties of each pulsar:  the pulsar's name, J2000 right ascension
and declination from the timing solution and the corresponding Galactic
coordinates.  This table also contains discovery information: the beam
in which the pulsar was detected, the radial angular distance of the pulsar
from the beam centre and the S/N of the pulse profile in the discovery
observation.  The median flux density averaged over all the
observations included in the timing solution and the pulse widths at 50
per cent and 10 per cent of the peak of the mean pulse profile are also
included (the 10 per cent width is only measured for pulsars with high
S/N).

Table~\ref{tb:per} contains the solar-system barycentric pulse period
$P$, period derivative $\dot P$, the epoch to which the period refers to, the number
of TOAs used in the timing solution, the MJD range covered by the
timing observations, the final rms value for the timing residuals and
the pulsar's DM. A non-deterministic period second derivative has also been fitted to a
few pulsars, indicated by asterisks in Table~\ref{tb:per}, which show 
significant amounts of timing noise.  

Data sets for each pulsar have been folded at twice and three times the
tabulated period in order to confirm that this represents the
fundamental period for the pulsar.  The TOAs range from
October 1997 to March 2002 with the shortest data span being 1.0
years and the longest 4.2 years. 

This paper includes data for four binary pulsars which have previously
been published:  PSRs~J1740$-$3052 \cite{sml+01}, J1810$-$2005,
J1904$+$0412 \cite{clm+01} and J1811$-$1736 \cite{lcm+00}.  For
completeness, their binary parameters are summarised in
Table~\ref{tb:binpar}.  PSR~J1837$-$0604, a young energetic pulsar,
has been discussed in D'Amico et~al.~(2001)\nocite{dkm+01} and between
MJDs 51305 and 51894, a glitch occurred in PSR~1806$-$2125 with a 
magnitude $\sim$2.5 times greater than any previously observed event \cite{hlj+02}.

For each pulsar, the profiles used in the determination of the TOAs
for the timing analysis were averaged to form mean pulse profiles
at 20\,cm (Fig.~\ref{fg:prf}).  For each profile, the pulsar's name,
pulse period and DM are given along with the effective time resolution
of the profile, and  pulsars timed at Parkes or Arecibo are indicated.
There are several small features in the baselines of some pulsar profiles
which may be other pulse components or may be radio frequency interference;  higher S/N
observations are required to resolve this issue.

\begin{table*}
\begin{minipage}{150mm}
\caption{Binary pulsar parameters. Data are taken from Stairs et al.
(2001) for J1740$-$3052, Camilo et al. (2001) for J1810$-$2005 and
J1904+0412 and Lyne et al. (2000) for J1811$-$1736.}
\begin{tabular}{lllll}\hline
                              & J1740$-$3052  & J1810$-$2005 & J1811$-$1736 & J1904$+$0412 \\\hline
Orbital period (d)      & 231.02965(3) & 15.0120197(9)& 18.779168(4)& 14.934263(2)\\
Projected semi-major axis (s) & 756.9087(4)  & 11.97791(8)  & 34.7830(8)  & 9.6348(1) \\
Eccentricity      	      & 0.5788720(4) & 0.000025(13) & 0.82802(2)& 0.00022(2)\\
Longitude of periastron(deg)  & 178.64613(6) & \ldots       & 127.661(2)& 350(6)\\
Epoch of ascending node (MJD) & 51353.51233(3)$^*$ & 51198.92979(2) &  51044.03702(3)$^*$ &51449.45(25) \\\hline
\end{tabular}   
\label{tb:binpar}
\\$^*$ epoch of periastron
\end{minipage}
\end{table*}

\section{DISCUSSION}

Table~\ref{tb:deriv} lists derived parameters for the 120 pulsars.  The
first column contains the pulsar's name, followed by $\log_{10}$ of the
characteristic age, $\tau_c = P/(2\dot{P})$ in years, the surface
dipole magnetic field strength, $B_s = 3.2\times
10^{19}(P\dot{P})^{1/2}$ in gauss and the rate of loss of rotational
energy, $\dot{E} = 4\pi^2 I\dot{P}P^{-3}$ in erg\,s$^{-1}$, where a
neutron star with moment of inertia $I = 10^{45}$\,g\,cm$^2$ is
assumed.  The next two columns give the pulsar distance, $d$, computed
from the DM assuming the Taylor \& Cordes (1993) \nocite{tc93} model
for the Galactic distribution of free electrons, together with the
corresponding Galactic $z$-height.  The uncertainty in distance from
this model is generally far greater than the precision with which the
distance is quoted.  The final column in Table~\ref{tb:deriv} gives the
radio luminosity $L_{1400} \equiv S_{1400} d^2$.

A $P$--$\dot{P}$ diagram is shown in Fig.~\ref{fg:ppdotdiagram}.  The
pulsars published here and in Papers~I and III are overlaid on
previously known pulsars.  Anomalous X-ray pulsars and soft
$\gamma$-ray repeaters are included along with lines of constant
surface magnetic field strength, a pulsar emission `death line' 
and a theoretical boundary separating
radio loud and radio quiet pulsars.

\subsection{The sample of 370 pulsars}

A full population analysis will follow the completion of the survey.
Here, we briefly highlight some of the salient features of the sample
of pulsars presented in this paper and in Papers~I and III.  To compare
the Parkes multibeam pulsars with previously known pulsars, Paper~I
displayed histograms of period, DM and flux density.  We continue this
comparison with histograms of pulse widths, ages, magnetic fields and
rate of loss of rotational energy for the 370 pulsars
(Fig.~\ref{fg:hist}).  There is a clear similarity between the measured
widths of the profiles of these pulsars and of previously known
pulsars (Fig.~\ref{fg:hist}a).  The characteristic ages for our sample are lower than
previously measured characteristic ages (Fig.~\ref{fg:hist}b).  
The observed age difference
likely arises because we search along the plane of the Galaxy, where
pulsars are born and where older pulsars have drifted from.  This is
also indicated by the results of the Swinburne intermediate latitude
survey which uses the same observing system and has discovered many
intermediate age to old pulsars \cite{ebs+01}.  Lorimer et~al.~(1995)
\nocite{lylg95} suggest that young pulsars have flatter radio spectra
than older pulsars.  Our high observing frequency may, therefore, be an
additional reason for the discovery of so many young pulsars.  However,
this difference in spectral indices is not confirmed by Malofeev~(1996)
\nocite{mal96} or Maron et~al.~(2000). \nocite{mkkw00b}

The possibility of a relationship between characteristic age and
luminosity was also studied; the results are shown in
Fig.~\ref{fg:lumPlots}a which is consistent with a constant radio
luminosity for ages $\sim 40$\,kyr--20\,Myr, although there seems to
be a significant reduction in luminosity for even older ages. 
The surface magnetic field histogram (Fig.~\ref{fg:hist}c) 
displays two features: the lack of pulsars with low magnetic
fields (reflecting the absence of millisecond and moderately recycled
pulsars in our sample) and the relatively large number of high magnetic
field pulsars, due to an excess of long period pulsars.  A graph 
of surface magnetic field versus luminosity
(Fig.~\ref{fg:lumPlots}b) is suggestive of the possibility that high
magnetic field pulsars have a higher luminosity than their lower-field
counterparts. A plot of 1400-MHz luminosity versus $\dot{E}$ for
our sample (Fig.~\ref{fg:lumPlots}c) shows no indication of any
trend, however, these figures do not take any geometrical beaming
effects into account.

The histogram of the rate of loss of rotational energy
(Fig.~\ref{fg:hist}d) contains a
component representing highly energetic pulsars
($\dot{E}>10^{36}$\,erg\,s$^{-1}$).  Three (PSRs~J1809$-$1917,
J1828$-$1101 and J1837$-$0604) are both young ($10^4 < \tau_c <
10^5$\,yr) and energetic.  PSR~J1913+1011 is slightly older ($\tau_c =
170$\,kyr), but has similar parameters to PSR~B1951+32 in the supernova
remnant CTB~80.  These `Vela-like' pulsars are discussed in Paper~III.
\begin{figure*}
\centerline{\psfig{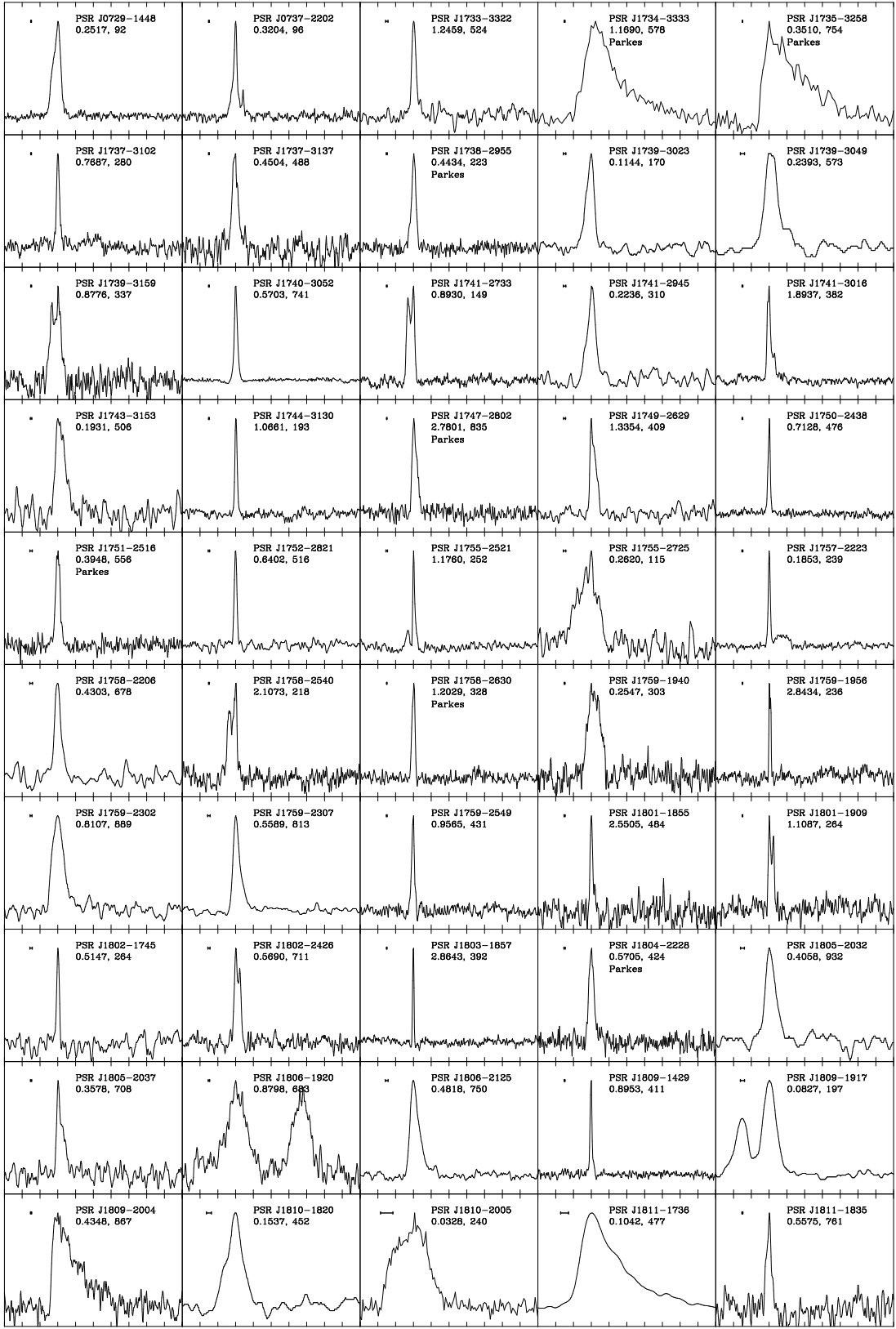}}
\caption{Mean pulse profiles for 120 pulsars discovered in the Parkes
multibeam survey. The highest point in the profile is placed at phase
0.3. For each profile, the pulsar's name, pulse period (s) and DM
(cm$^{-3}$\,pc) are given. The small horizontal bar to the left of the
pulse indicates the effective time resolution of the profile, including
the effects of interstellar dispersion.}
\label{fg:prf}
\end{figure*}
\addtocounter{figure}{-1}
\begin{figure*} 
\centerline{\psfig{file=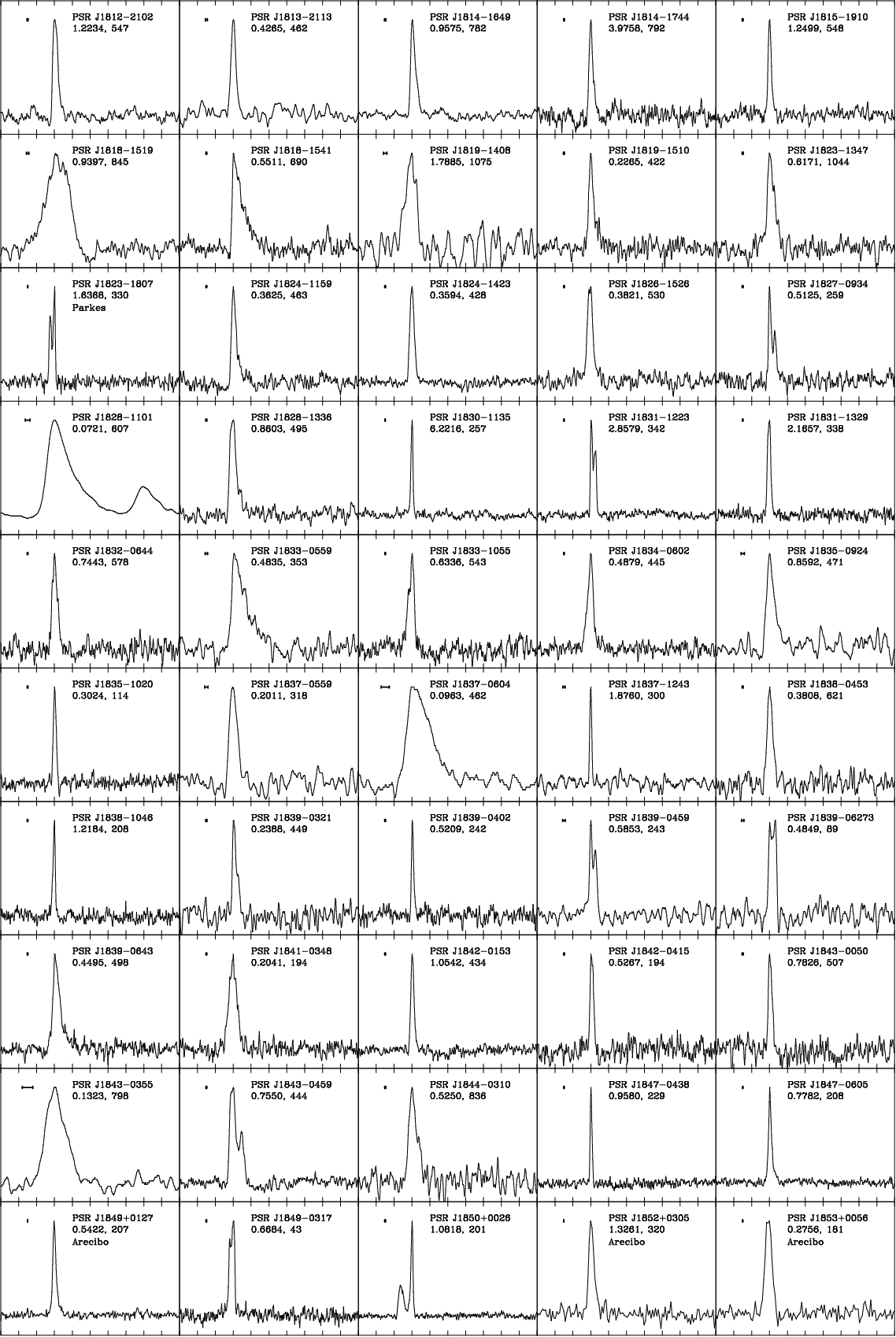,width=145mm}}
\caption{-- {\it continued}}
\end{figure*}
\addtocounter{figure}{-1}
\begin{figure*} 
\centerline{\psfig{file=mb1069fg2c.ps,width=145mm}}
\caption{-- {\it continued}}
\end{figure*}

\begin{figure*} 
\centerline{\psfig{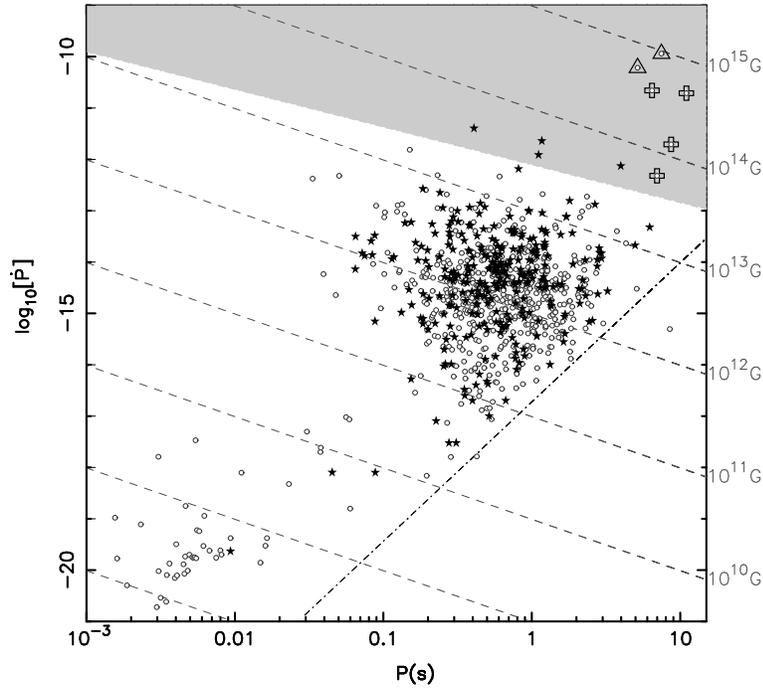}} 
\caption{$P$--$\dot{P}$ diagram for the 370 pulsars published in this
paper and in Papers~I and III (stars) overlaid on previously known
pulsars.  Lines of constant magnetic field are
shown as dashed lines. PSRs~J1119$-$6217, J1726$-$3530, J1814$-$1744 (Camilo
et~al.~2000) and J1734$-$3333 lie above the theoretical boundary
between radio loud and radio quiet pulsars (shown as the light shaded
region). The boundary of this region is given by equation 10 in Baring
\& Harding (2001).  Some of these pulsars are also close to the
anomalous X-ray pulsars (AXPs), indicated as open crosses and the
soft $\gamma$-ray repeaters (SGRs) shown as triangles.  The
placement of the AXPs and SGRs on this diagram assumes that they are
spinning down due to magnetic dipole radiation in a manner similar to
the radio pulsars.  The `death line' (dot-dashed line) is defined by $7\log B_s - 13\log P
= 78$ (Chen \& Ruderman 1993). }
\label{fg:ppdotdiagram}
\end{figure*}
\nocite{cr93a}

\begin{figure*}
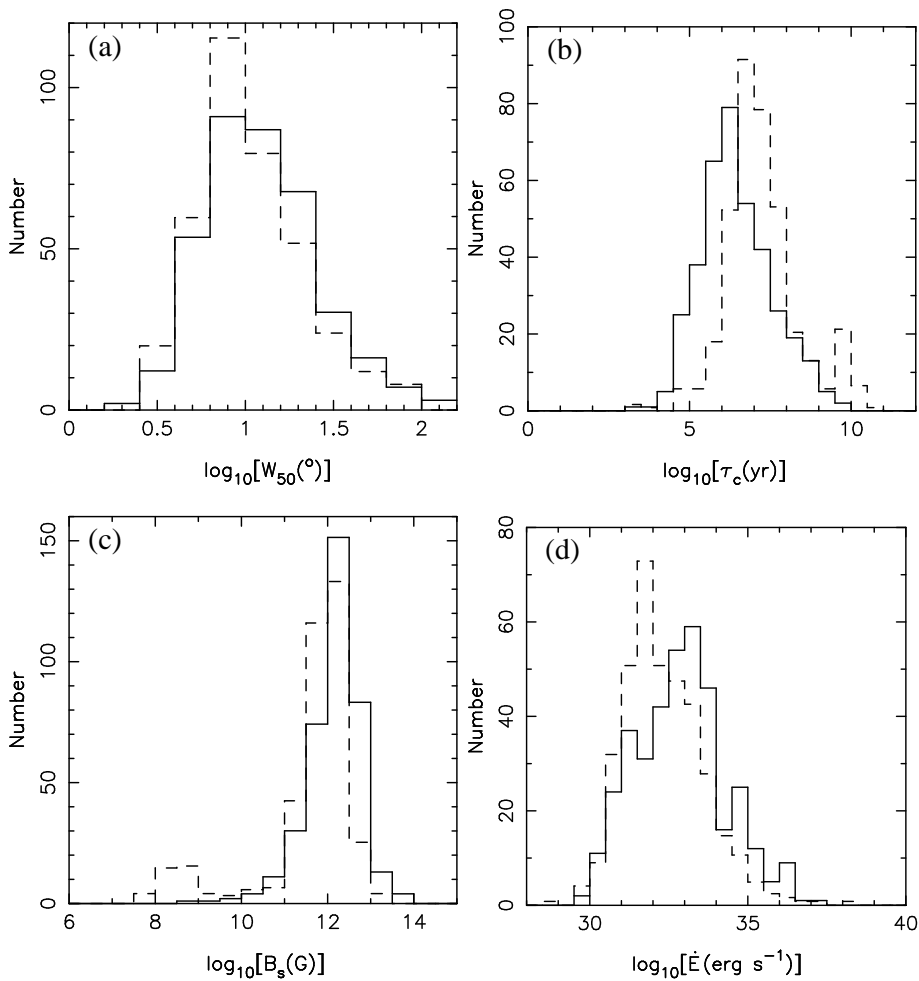
 
\begin{center}\begin{minipage}[t]{50mm}
\begin{center}\psfig{file=mb1069fg4a.ps,width=60mm} \end{center}
\end{minipage}
\begin{minipage}[t]{50mm}
\begin{center}\psfig{file=mb1069fg4b.ps,width=60mm} \end{center}
\end{minipage}
\begin{minipage}[t]{50mm}
\begin{center}\psfig{file=mb1069fg4c.ps,width=60mm} \end{center}
\end{minipage}
\begin{minipage}[t]{50mm}
\begin{center}\psfig{file=mb1069fg4d.ps,width=60mm} \end{center}
\end{minipage}\end{center}
\caption{(a) Distribution in pulse width at 50\% peak height for 370
multibeam pulsars (solid line) and for previously known pulsars (dashed
line).  For the latter, the vertical scale has been adjusted to
equalise the number of pulsars in the two distributions. Panels (b),
(c) and (d) similarly display distributions of characteristic age,
surface magnetic field strength and rate of energy loss respectively. }
\label{fg:hist}
\end{figure*}                                                  

\begin{figure*}
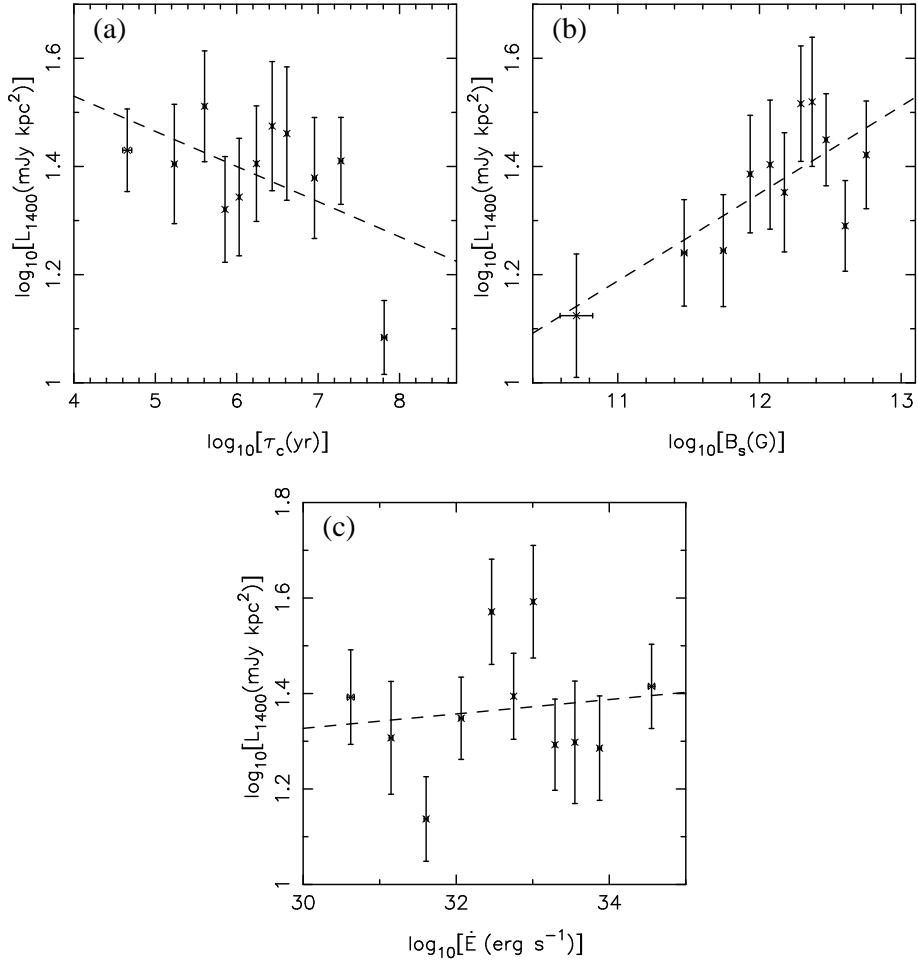

\begin{center}\begin{minipage}[t]{60mm}
\begin{center}\psfig{file=mb1069fg5a.ps,width=60mm,angle=-90} \end{center}
\end{minipage}
\begin{minipage}[t]{60mm}
\begin{center}\psfig{file=mb1069fg5b.ps,width=60mm,angle=-90} \end{center}
\end{minipage}
\begin{minipage}[t]{60mm}
\begin{center}\psfig{file=mb1069fg5c.ps,width=60mm,angle=-90} \end{center}
\end{minipage}
\end{center}
\caption{ (a) Radio luminosity versus characteristic age ($\propto
P\dot{P}^{-1}$), (b) luminosity versus surface magnetic field
strength ($\propto [P\dot{P}]^{1/2}$) and (c) luminosity versus rate
of loss of rotational energy ($\propto \dot{P}P^{-3}$) for 330 pulsars.  Each point
represents the average of 30 pulsars binned along the horizontal axis.  
The error bars for each point reflect the statistical
scatter within each bin of 30 pulsars, and do not include other errors
such as that due to the distance model uncertainty.  The dashed lines
represent a straight line fit to the data with a gradient of $(-0.06
\pm 0.04)$ for (a), $(0.16 \pm 0.05)$ for (b) and $(0.02 \pm 0.04)$ for
(c).} 
\label{fg:lumPlots}
\end{figure*}

The average pulse profiles (Fig.~\ref{fg:prf}) are diverse: the majority
contain a single component, $\sim$~20\% contain multiple components and
three (PSRs~J1806$-$1920, J1828$-$1101 and J1913$+$0832) clearly have
interpulses.  Eight (PSRs~J1734$-$3333, J1735$-$3258, J1809$-$2004,
J1811$-$1736, J1818$-$1541, J1828$-$1101, J1833$-$0559 and
J1837$-$0604) have features suggestive
of scattering tails.  Although in general pulsars showing evidence of
scattering also have high DM, we find no simple relationship between DM
and the width of the scattering tail (see Paper~I).

\subsection{PSR~J1734$-$3333}

Camilo et~al.~(2000) \nocite{ckl+00} discuss PSR~J1814$-$1744, whose
inferred surface magnetic field strength, $B_s = 5.5\times10^{13}$\,G,
is the largest of any known radio pulsar.  PSR~J1734$-$3333 has a
similar field strength of $5.2\times10^{13}$\,G (and has a very small
age, $\tau_c = 8$\,kyr).  Both these values were calculated using the
standard expression\footnote{Note that occasionally in the literature
the magnetic field is calculated with an additional factor of two.  In
that case the field represents the maximum surface dipole field, as
opposed to the value at the equator.} $B_s =
3.2\times10^{19}(P\dot{P})^{1/2}$\,G.  Taking these fields at face
value, these pulsars thus have surface magnetic field strengths that
are larger than the `quantum critical field', $B_c = 4.4\times
10^{13}$\,G, at which the cyclotron energy equals the electron
rest-mass energy.  Baring \& Harding~(2001) \nocite{bh01} propose that
photon splitting may inhibit pair creation at these high magnetic
fields and therefore hypothesize the existence of a boundary separating
radio-loud and radio-quiet pulsars.  This boundary is indicated on
Fig.~\ref{fg:ppdotdiagram} assuming that the emission occurs from the
surface of the neutron star and that the photons propagate parallel to
the local field.  PSRs~J1734$-$3333 and J1814$-$1744 lie in the
predicted radio quiescence region. However, as detailed in Baring \&
Harding~(2001) \nocite{bh01} the boundary moves up the $P$--$\dot{P}$
diagram for emission away from the neutron star's surface.  For an
emission radius of 1.5 times the neutron star's radius, the boundary
lies above all the known radio pulsars.  We note that given the trend
toward pulse narrowing with increasing spin period \cite{gou94}, the
fact that no radio counterparts of AXPs, which also lie above this
boundary, have been detected, may be due to their radio beams not
intersecting our line of sight \cite{gsg01}.

The success of this survey at finding new pulsars with fields close to
and above the critical field will help improve our understanding of the
delineation between radio-loud and radio-quiet pulsars.

\subsection{PSR~J1830$-$1135} 

After the discovery of PSR~J2144$-$3933, a radio pulsar with an 8.5-s
period \cite{ymj99}, there has been much interest in detecting
long-period pulsars.  PSR~J1830$-$1135, with a period of 6-s, is the
second slowest known radio pulsar.  It lies on the $P$--$\dot{P}$
diagram at the extreme right-hand edge of the `normal' pulsars, but due
to a relatively high surface magnetic field of $\sim 10^{13}$\,Gauss,
lies well above the `death line' (Fig.~\ref{fg:ppdotdiagram}).  Although
the magnetic field is below the critical field, PSR~J1830$-$1135 (and
PSR~J1814$-$1744) lies roughly in the neighbourhood of some AXPs (shown as
open crosses in Fig.~\ref{fg:ppdotdiagram}).  Pivovaroff, Kaspi \&
Camilo~(2000) \nocite{pkc00} infer, using PSR~J1814$-$1744, that the
X-ray emission from AXPs depends upon more than the implied surface
magnetic field strength.  The reasons for this are, however, not
well understood.

Among the 370 published Parkes multibeam pulsars, only PSR~J1830$-$1135
has a period greater than 6 seconds.  Although the search software
limits candidates to periods below 5 seconds (see Paper~I),  strong
long-period pulsars with narrow pulse profiles, such as
PSR~J1830$-$1135, may be detected from their second or third
harmonics.  The probability of detecting a pulsar is also dependent
upon the width of the pulsar beam, which decreases with increasing
period (e.g.  Rankin~1983 \nocite{ran83a} and
Gould~1994\nocite{gou94}).  Although determination of the physical beam
size requires polarisation information, the small angular pulse width
of $\sim 4\degr$ for PSR~J1830$-$1135 is compatible with current
understanding.

\subsection{Conclusion}

The sample of 370 pulsars reported here and in Papers~I and III
represents approximately half of the total number of new pulsars that
we expect to discover on completion of this survey.  Details of
the remaining pulsars will be published in due course as the data from
the timing observations become available.  Once complete, the survey
will provide a large and relatively unbiased dataset which may be used
to study many aspects of pulsar physics.

\section*{Acknowledgements} 

We gratefully acknowledge the technical assistance provided by George
Loone, Tim Ikin, Mike Kesteven, Mark Leach and the staff at the Parkes
and Jodrell Bank Observatories toward the development of the Parkes
multibeam pulsar system.  We thank R.~Bhat, D.~Lorimer, M.~McLaughlin
and P.~Freire for assistance with observations. DJM and GH acknowledge the receipt of
post-graduate studentships from the UK Particle Physics and Astronomy
Research Council. IHS received support from NSERC and Jansky
postdoctoral fellowships.  FC is supported by NASA grants
NAG~5-9095 and NAG~5-9950.  VMK is an Alfred P. Sloan Research Fellow and was
supported in part by a US National Science Foundation (NSF) Career
Award (AST-9875897) and by a Natural Sciences and Engineering Research
Council of Canada grant (RGPIN 228738-00). The Parkes radio telescope
is part of the Australia Telescope which is funded by the Commonwealth
of Australia for operation as a National Facility managed by CSIRO. The
Arecibo Observatory is part of the National Astronomy and Ionosphere
Center, which is operated by Cornell University under a cooperative
agreement with the US NSF.

\bibliography{journals,modrefs,psrrefs,crossrefs}
\bibliographystyle{mn}

\newpage

\input{mb1069tb4a.tex}
\addtocounter{table}{-1}
\input{mb1069tb4b.tex}

\addtocounter{table}{-1}
\input{mb1069tb4c.tex}

\end{document}

%% file: mb1069tb1a.tex
\begin{table*}
\begin{minipage}{150mm}
\caption{Positions, flux densities and pulse widths for 120
higher declination pulsars discovered in the Parkes multibeam pulsar
survey.  `A' indicates pulsars that have been timed at Arecibo and `P'
denotes pulsars timed using the Parkes telescope.  All other pulsars
were timed using the Lovell telescope at Jodrell Bank. Radial angular distances
are given in units of beam radii.  PSRs~J1841$-$0348, J1842$-$0415 and J1844$-$0310
were discovered independently by Lorimer et~al.~(2000).}
\label{tb:posn}
\begin{tabular}{lllrrccrllr}
\hline 
\multicolumn{1}{c}{PSR J} & R.A. (J2000) & Dec. (J2000) & 
\multicolumn{1}{c}{$l$} & \multicolumn{1}{c}{$b$} & Beam & Radial & 
\multicolumn{1}{c}{$S/N$} & \multicolumn{1}{c}{$S_{1400}$} & \multicolumn{1}{c} {$W_{50}$} & 
\multicolumn{1}{c}{$W_{10}$} \\ & (h~~~m~~~s) & (~\degr ~~~\arcmin ~~~\arcsec) & 
\multicolumn{1}{c}{(\degr)} & \multicolumn{1}{c}{(\degr)} &   &
Dist. &  & 
\multicolumn{1}{c}{(mJy)} & \multicolumn{1}{c}{(ms)} & \multicolumn{1}{c}{(ms)} \\ 
\hline 
0729$-$1448 		&	 07:29:16.45(2) &	 $-$14:48:36.8(8) &	 230.39 &	 1.42 		&	 12 &	 0.27 &	 49.1 &	 0.7(1) &	 12.5 &	 22 \\ 
0737$-$2202 		&	 07:37:44.071(6) &	 $-$22:02:05.3(1) &	 237.69 &	 $-$0.32 	&	 9 &	 0.87 &	 23.6 &	 0.47(9) &	 6.3 &	 20 \\ 
1733$-$3322 		&	 17:33:55.21(8) &	 $-$33:22:03(4) &	 354.92 &	 $-$0.24 	&	 1 &	 0.63 &	 61.2 &	 0.8(2) &	 34.3 &	 -- \\ 
1734$-$3333$^{\rm P}$ 	&	 17:34:26.6(5) &	 $-$33:33:22(29) &	 354.81 &	 $-$0.44 	&	 7 &	 0.48 &	 20.0 &	 0.5(1) &	 164.1 &	 -- \\ 
1735$-$3258$^{\rm P}$ 	&	 17:35:56.9(3) &	 $-$32:58:19(19) &	 355.48 &	 $-$0.38 	&	 5 &	 0.92 &	 10.1 &	 0.46(9) &	 187.5 &	 -- \\ 
& & & & & & & & & & \\ 
1737$-$3102 		&	 17:37:33.73(4) &	 $-$31:02:01(4) &	 357.30 &	 0.37 		&	 9 &	 0.57 &	 39.8 &	 0.6(1) &	 15.0 &	 36 \\ 
1737$-$3137 		&	 17:37:04.29(4) &	 $-$31:37:21(3) &	 356.74 &	 0.14 		&	 10 &	 1.23 &	 20.2 &	 0.8(2) &	 16.4 &	 -- \\ 
1738$-$2955$^{\rm P}$ 	&	 17:38:52.24(10) &	 $-$29:55:51(10) &	 358.38 &	 0.72 		&	 2 & 0.89 &	 16.2 &	 0.29(6) &	 23.4 &	 62 \\ 
1739$-$3023 		&	 17:39:39.80(2) &	 $-$30:23:12(2) &	 358.09 &	 0.34 		&	 1 &	 0.66 &	 49.1 &	 1.0(2) &	 6.1 &	 11 \\ 
1739$-$3049 		&	 17:39:23.22(6) &	 $-$30:49:40(5) &	 357.68 &	 0.15 		&	 4 &	 0.86 &	 17.6 &	 0.5(1) &	 18.8 &	 -- \\ 
& & & & & & & & & & \\ 
1739$-$3159 		&	 17:39:48.68(8) &	 $-$31:59:49(9) &	 356.74 &	 $-$0.55 	&	 1 &	 1.06 &	 33.0 &	 1.0(2) &	 53.5 &	 -- \\ 
1740$-$3052 		&	 17:40:50.031(5) &	 $-$30:52:04.1(3) &	 357.81 &	 $-$0.13 	&	 7 &	 1.19 &	 26.1 &	 0.7(2) &	 8.9 &	 27 \\ 
1741$-$2733 		&	 17:41:01.34(4) &	$-$27:33:51(7) &	 0.64 &	 	 1.58 		&	 3 &	 0.61 &	 77.2 &	 1.1(2) &	 42.2 &	 55 \\ 
1741$-$2945 		&	 17:41:14.47(4) &	 $-$29:45:35(5) &	 358.80 &	 0.38 		&	 12 &	 0.36 &	 15.9 &	 0.6(1) &	 10.7 &	 -- \\ 
1741$-$3016 		&	 17:41:07.04(6) &	 $-$30:16:31(10) &	 358.35 &	 0.13 		&	 6 &	 1.26 &	 38.7 &	 2.3(5) &	 39.8 &	 111 \\ 
& & & & & & & & & & \\ 
1743$-$3153 		&	 17:43:15.57(1) &	 $-$31:53:05(2) &	 357.22 &	 $-$1.11 	&	 6 &	 1.01 &	 10.9 &	 0.5(1) &	 11.3 &	 -- \\ 
1744$-$3130 		&	 17:44:05.68(1) &	 $-$31:30:04(3) &	 357.64 &	 $-$1.06 	&	 12 &	 0.41 &	 62.1 &	 0.7(1) &	 16.0 &	 29 \\ 
1747$-$2802$^{\rm P}$ 	&	 17:47:26.6(2) &	 $-$28:02:37(33) &	 0.97 &	 0.12 			&	 6 &	 1.03 &	 10.2 &	 0.5(1) &	 27.3 &	 -- \\ 
1749$-$2629 		&	 17:49:11.28(5) &	 $-$26:29:10(19) &	 2.50 &	 0.59 			&	 7 &	 0.35 &	 52.6 &	 0.7(2) &	 46.4 &	 -- \\ 
1750$-$2438 		&	 17:50:59.787(9) &	 $-$24:38:58(7) &	 4.29 &	 1.19 			&	 13 &	 0.90 &	 41.2 &	 0.5(1) &	 9.6 &	 19 \\ 
& & & & & & & & & & \\ 
1751$-$2516$^{\rm P}$ 	&	 17:51:52.63(6) &	 $-$25:16:43(26) &	 3.85 &	 0.69 			&	 1 &	 0.72 &	 7.0 &	 0.22(4) &	 27.3 &	 -- \\ 
1752$-$2821 		&	 17:52:24.55(3) &	 $-$28:21:10(9) &	 1.27 &	 $-$0.98 		&	 6 &	 0.59 &	 29.4 &	 0.32(6) &	 11.5 &	 20 \\ 
1755$-$2521 		&	 17:55:59.71(9) &	 $-$25:21:27(48) &	 4.19 &	 $-$0.19 			&	 7 &	 1.49 &	 17.3 &	 0.7(2) &	 13.4 &	 33 \\ 
1755$-$2725 		&	 17:55:41.88(9) &	 $-$27:25:45(24) &	 2.43 &	 $-$1.14 		&	 7 &	 0.97 &	 18.4 &	 0.5(1) &	 23.8 &	 -- \\ 
1757$-$2223 		&	 17:57:50.772(4) &	 $-$22:23:49(4) &	 7.03 &	 0.97 			&	 3 &	 0.84 &	 44.8 &	 1.1(2) &	 2.3 &	 5 \\ 
& & & & & & & & & & \\ 
1758$-$2206 		&	 17:58:44.45(3) &	 $-$22:06:45(19) &	 7.38 &	 0.93 			&	 8 &	 0.51 &	 25.4 &	 0.41(8) &	 16.6 &	 -- \\ 
1758$-$2540 		&	 17:58:31.94(9) &	 $-$25:40:49(46) &	 4.26 &	 $-$0.81 		&	 6 &	 0.62 &	 35.1 &	 0.65(1) &	 120.0 &	 -- \\ 
1758$-$2630$^{\rm P}$ 	&	 17:58:34.3(1) &	 $-$26:30:10(21) &	 3.55 &	 $-$1.23 		&	 9 &	 0.14 &	 16.1 &	 0.41(8) &	 15.6 &	 -- \\ 
1759$-$1940 		&	 17:59:57.04(2) &	 $-$19:40:29(5) &	 9.63 &	 1.90 			&	 6 &	 0.98 &	 39.1 &	 0.9(2) &	 19.6 &	 -- \\ 
1759$-$1956 		&	 17:59:35.42(4) &	 $-$19:56:08(15) &	 9.37 &	 1.84 &	 4 &	 0.68 &	 49.8 &	 0.41(8) &	 37.9 &	 -- \\ 
& & & & & & & & & & \\ 
1759$-$2302 		&	 17:59:49.23(9) &	 $-$23:02(3) &	 6.70 &	 0.26 &	 10 &	 1.04 &	 34.5 &	 1.3(3) &	 61.5 &	 -- \\ 
1759$-$2307 		&	 17:59:30.95(2) &	 $-$23:07:17(46) &	 6.59 &	 0.27 &	 9 &	 0.47 &	 24.1 &	 0.7(1) &	 22.4 &	 52 \\ 
1759$-$2549 		&	 17:59:35.12(9) &	 $-$25:49:07(34) &	 4.26 &	 $-$1.08 &	 5 &	 1.30 &	 11.6 &	 0.6(1) &	 15.6 &	 -- \\ 
1801$-$1855 		&	 18:01:22.3(3) &	 $-$18:55:49(49) &	 10.45 &	 1.98 &	 12 &	 0.52 &	 26.2 &	 0.47(9) &	 41.9 &	 -- \\ 
1801$-$1909 		&	 18:01:46.69(3) &	 $-$19:09:36(6) &	 10.30 &	 1.78 &	 9 &	 0.97 &	 26.6 &	 0.5(1) &	 40.0 &	 -- \\ 
& & & & & & & & & & \\ 
1802$-$1745 		&	 18:02:14.85(3) &	 $-$17:45:17(7) &	 11.57 &	 2.38 &	 10 &	 0.57 &	 19.5 &	 0.21(4) &	 8.8 &	 -- \\ 
1802$-$2426 		&	 18:02:03.10(4) &	 $-$24:26:43(30) &	 5.73 &	 $-$0.89 &	 10 &	 1.14 &	 13.8 &	 0.6(1) &	 22.2 &	 -- \\ 
1803$-$1857 		&	 18:03:59.06(3) &	 $-$18:57:19(8) &	 10.73 &	 1.43 &	 2 &	 0.98 &	 33.3 &	 0.40(8) &	 25.6 &	 43 \\ 
1804$-$2228$^{\rm P}$ 	&	 18:04:28.19(9) &	 $-$22:28:18(55) &	 7.72 &	 $-$0.40 &	 5 &	 0.04 &	 10.1 &	 0.20(4) &	 27.3 &	 -- \\ 
1805$-$2032 		&	 18:05:37.09(6) &	 $-$20:32:51(21) &	 9.53 &	 0.31 &	 2 &	 0.81 &	 21.4 &	 0.7(1) &	 26.3 &	 -- \\ 
& & & & & & & & & & \\ 
1805$-$2037 		&	 18:05:28.19(3) &	 $-$20:37:16(11) &	 9.45 &	 0.31 &	 5 &	 1.16 &	 10.6 &	 0.34(7) &	 13.2 &	 -- \\ 
1806$-$1920 		&	 18:06:06.6(2) &	 $-$19:20:23(40) &	 10.63 &	 0.80 &	 11 &	 1.14 &	 15.6 &	 1.9(4) &	 400.0 &	 -- \\ 
1806$-$2125 		&	 18:06:19.59(9) &	 $-$21:25:41(24) &	 8.86 &	 $-$0.25 &	 3 &	 1.17 &	 22.6 &	 1.1(2) &	 26.4 &	 55 \\ 
1809$-$1429 		&	 18:09:45.510(8) &	 $-$14:29:25(1) &	 15.31 &	 2.39 &	 12 &	 1.52 &	 20.8 &	 0.6(1) &	 11.0 &	 27 \\ 
1809$-$1917 		&	 18:09:43.147(6) &	 $-$19:17:38(1) &	 11.09 &	 0.08 &	 2 &	 1.06 &	 33.8 &	 2.5(5) &	 18.0 &	 26 \\ 
& & & & & & & & & & \\ 
1809$-$2004 		&	 18:09:15.9(2) &	 $-$20:04:12(57) &	 10.36 &	 $-$0.20 &	 10 &	 0.35 &	 24.2 &	 0.9(2) &	 46.3 &	 -- \\ 
1810$-$1820 		&	 18:10:55.52(3) &	 $-$18:20:39(6) &	 12.06 &	 0.29 &	 5 &	 0.78 &	 19.0 &	 0.7(2) &	 13.6 &	 31 \\ 
1810$-$2005 		&	 18:10:58.988(2) &	 $-$20:05:08.3(6) &	 10.54 &	 $-$0.56 &	 4 &	 0.77 &	 46.2 &	 2.0(4) &	 7.2 &	 11 \\ 
1811$-$1736 		&	 18:11:55.01(1) &	 $-$17:36:36.9(13) &	 12.82 &	 0.44 &	 10 &	 0.62 &	 21.3 &	 1.3(3) &	 23.4 &	 63 \\ 
1811$-$1835 		&	 18:11:29.72(9) &	 $-$18:35:44(13) &	 11.91 &	 0.05 &	 1 &	 0.44 &	 25.4 &	 0.42(8) &	 16.7 &	 -- \\ 
\hline
\end{tabular}
\end{minipage}
\end{table*}

%% file: mb1069tb1b.tex
\begin{table*}
\begin{minipage}{150mm}
\caption{-- {\it continued}}
\begin{tabular}{lllrrccrllr}
\hline 
\multicolumn{1}{c}{PSR J} & R.A. (J2000) & Dec. (J2000) & 
\multicolumn{1}{c}{$l$} & \multicolumn{1}{c}{$b$} & Beam & Radial & 
\multicolumn{1}{c}{$S/N$} & \multicolumn{1}{c}{$S_{1400}$} & \multicolumn{1}{c} {$W_{50}$} & 
\multicolumn{1}{c}{$W_{10}$} \\ & (h~~~m~~~s) & (~\degr ~~~\arcmin ~~~\arcsec) & 
\multicolumn{1}{c}{(\degr)} & \multicolumn{1}{c}{(\degr)} &   & Dist. &  & 
\multicolumn{1}{c}{(mJy)} & \multicolumn{1}{c}{(ms)} & \multicolumn{1}{c}{(ms)} \\ 
\hline 
1812$-$2102 &	 18:12:20.93(6) &	 $-$21:02:36(15) &	 9.86 &	 $-$1.30 &	 12 &	0.61 &	 58.3 &	 1.4(3) &	 39.4 &	 66 \\ 
1813$-$2113 &	 18:13:39.87(3) &	 $-$21:13:00(7) &	 9.85 &	 $-$1.66 &	 11 &	 0.72 &	 46.0 &	 0.6(1) &	 13.1 &	 -- \\ 
1814$-$1649 &	 18:14:37.35(4) &	 $-$16:49:28(5) &	 13.82 &	 0.25 &	 11 &	 0.83 &	 45.0 &	 1.1(2) &	 28.0 &	 58 \\ 
1814$-$1744 &	 18:14:43.0(2) &	 $-$17:44:33(33) &	 13.02 &	 $-$0.21 &	 1 &	 1.40 &	 22.1 &	 0.7(1) &	 92.0 &	 -- \\ 
1815$-$1910 &	 18:15:03.08(4) &	 $-$19:10:00(8) &	 11.81 &	 $-$0.96 &	 3 &	 0.43 &	 16.0 &	 0.32(6) &	 25.4 &	 52 \\ 
& & & & & & & & & & \\ 
1818$-$1519 &	 18:18:14.6(3) &	 $-$15:19:43(36) &	 15.55 &	 0.19 &	 9 &	 0.54 &	 66.2 &	 2.1(4) &	 130.4 &	 -- \\ 
1818$-$1541 &	 18:18:37.52(3) &	 $-$15:41:45(5) &	 15.27 &	 $-$0.06 &	 8 &	 0.88 &	 26.2 &	 1.0(2) &	 27.2 &	 -- \\ 
1819$-$1408 &	 18:19:56.8(5) &	 $-$14:08:02(50) &	 16.80 &	 0.39 &	 5 &	 0.19 &	 27.4 &	 0.5(1) &	 116.8 &	 -- \\ 
1819$-$1510 &	 18:19:53.691(8) &	 $-$15:10:21(1) &	 15.88 &	 $-$0.09 &	 7 &	 0.66 &	 36.8 &	 0.6(1) &	 6.9 &	 -- \\ 
1823$-$1347 &	 18:23:24.25(8) &	 $-$13:47:54(9) &	 17.49 &	 $-$0.19 &	 4 &	 1.01 &	 14.5 &	 0.41(8) &	 34.6 &	 -- \\ 
& & & & & & & & & & \\ 
1823$-$1807$^{\rm P}$ &	 18:23:09.68(7) &	 $-$18:07:33(8) &	 13.64 &	 $-$2.16 &	 7 &	 0.67 &	 14.6 &	 0.39(8) &	 34.6 &	 -- \\ 
1824$-$1159 &	 18:24:56.15(1) &	 $-$11:59:53(1) &	 19.25 &	 0.32 &	 6 &	 0.87 &	 25.9 &	 0.7(1) &	 10.8 &	 -- \\ 
1824$-$1423 &	 18:24:57.387(7) &	 $-$14:23:05(1) &	 17.15 &	 $-$0.80 &	 4 &	 0.92 &	 49.0 &	 0.8(2) &	 9.9 &	 18 \\ 
1826$-$1526 &	 18:26:12.60(3) &	 $-$15:26:03(5) &	 16.36 &	 $-$1.55 &	 4 &	 0.76 &	 22.4 &	 0.46(9) &	 13.4 &	 -- \\ 
1827$-$0934 &	 18:27:45.88(2) &	 $-$09:34:16(1) &	 21.72 &	 0.84 &	 9 &	 0.45 &	 26.5 &	 0.29(6) &	 19.5 &	 -- \\ 
& & & & & & & & & & \\ 
1828$-$1101 &	 18:28:18.859(9) &	 $-$11:01:50.9(8) &	 20.49 &	 0.04 &	 13 &	 0.74 &	 34.8 &	 2.9(6) &	 9.9 &	 49 \\ 
1828$-$1336 &	 18:28:42.85(4) &	 $-$13:36:45(5) &	 18.25 &	 $-$1.24 &	 6 &	 0.62 &	 36.5 &	 0.26(5) &	 34.6 &	 -- \\ 
1830$-$1135 &	 18:30:01.72(6) &	 $-$11:35:32(6) &	 20.19 &	 $-$0.59 &	 9 &	 0.36 &	 75.3 &	 1.1(2) &	 62.2 &	 171 \\ 
1831$-$1223 &	 18:31:12.8(1) &	 $-$12:23:31(11) &	 19.62 &	 $-$1.22 &	 10 &	 0.30 &	 77.5 &	 1.2(2) &	 98.0 &	 122 \\ 
1831$-$1329 &	 18:31:55.96(6) &	 $-$13:29:56(8) &	 18.72 &	 $-$1.88 &	 13 &	 0.87 &	 40.4 &	 0.5(1) &	 47.1 &	 82 \\ 
& & & & & & & & & & \\ 
1832$-$0644 &	 18:32:42.67(5) &	 $-$06:44:02(3) &	 24.81 &	 1.07 &	 4 &	 0.69 &	 49.4 &	 0.7(1) &	 27.3 &	 -- \\ 
1833$-$0559 &	 18:33:07.60(5) &	 $-$05:59:25(5) &	 25.51 &	 1.32 &	 2 &	 0.95 &	 26.6 &	 0.6(1) &	 41.4 &	 -- \\ 
1833$-$1055 &	 18:33:58.40(3) &	 $-$10:55:31(3) &	 21.23 &	 $-$1.14 &	 6 &	 0.76 &	 38.4 &	 0.5(1) &	 24.2 &	 -- \\ 
1834$-$0602 &	 18:34:37.97(3) &	 $-$06:02:35(3) &	 25.64 &	 0.96 &	 12 &	 0.54 &	 34.2 &	 0.8(2) &	 16.8 &	 35 \\ 
1835$-$0924 &	 18:35:38.0(1) &	 $-$09:24:27(13) &	 22.77 &	 $-$0.80 &	 7 &	 0.90 &	 18.1 &	 0.5(1) &	 41.6 &	 -- \\ 
& & & & & & & & & & \\ 
1835$-$1020 &	 18:35:57.55(1) &	 $-$10:20:05(1) &	 21.98 &	 $-$1.30 &	 12 &	 0.58 &	 123.3 &	 1.9(4) &	 5.7 &	 10 \\ 
1837$-$0559 &	 18:37:23.66(1) &	 $-$05:59:28.3(9) &	 26.00 &	 0.38 &	 9 &	 0.85 &	 18.8 &	 0.5(1) &	 10.7 &	 -- \\ 
1837$-$0604 &	 18:37:43.8(1) &	 $-$06:04:52(7) &	 25.96 &	 0.26 &	 11 &	 0.80 &	 17.7 &	 0.7(2) &	 13.0 &	 30 \\ 
1837$-$1243 &	 18:37:09.41(3) &	 $-$12:43:56(3) &	 19.98 &	 $-$2.66 &	 10 &	 0.25 &	 26.1 &	 0.17(3) &	 24.4 &	 -- \\ 
1838$-$0453 &	 18:38:11.19(6) &	 $-$04:53:23(3) &	 27.07 &	 0.71 &	 2 &	 0.84 &	 18.4 &	 0.33(7) &	 12.7 &	 -- \\ 
& & & & & & & & & & \\ 
1838$-$1046 &	 18:38:26.48(4) &	 $-$10:46:57(4) &	 21.86 &	 $-$2.05 &	 2 &	 1.49 &	 14.3 &	 0.5(1) &	 16.6 &	 34 \\ 
1839$-$0321 &	 18:39:37.51(2) &	 $-$03:21:11(1) &	 28.60 &	 1.09 &	 6 &	 0.55 &	 15.7 &	 0.27(5) &	 5.4 &	 -- \\ 
1839$-$0402 &	 18:39:51.06(3) &	 $-$04:02:25(1) &	 28.02 &	 0.73 &	 5 &	 0.83 &	 23.0 &	 0.21(4) &	 7.0 &	 15 \\ 
1839$-$0459 &	 18:39:42.62(2) &	 $-$04:59:59(2) &	 27.15 &	 0.32 &	 7 &	 0.86 &	 19.8 &	 0.30(6) &	 11.7 &	 -- \\ 
1839$-$06273 &	 18:39:20.46(3) &	 $-$06:27:34(5) &	 25.81 &	 $-$0.27 &	 5 &	 0.58 &	 13.6 &	 0.29(6) &	 23.2 &	 -- \\ 
& & & & & & & & & & \\ 
1839$-$0643 &	 18:39:09.80(1) &	 $-$06:43:45(1) &	 25.55 &	 $-$0.35 &	 1 &	 1.10 &	 42.6 &	 1.4(3) &	 19.1 &	 -- \\ 
1841$-$0348 &	 18:41:38.68(5) &	 $-$03:48:43(2) &	 28.42 &	 0.44 &	 1 &	 0.51 &	 74.3 &	 1.4(3) &	 9.5 &	 18 \\ 
1842$-$0153 &	 18:42:57.87(1) &	 $-$01:53:26.7(6) &	 30.28 &	 1.02 &	 8 &	 0.83 &	 28.0 &	 0.6(1) &	 21.4 &	 -- \\ 
1842$-$0415 &	 18:42:11.29(1) &	 $-$04:15:38.2(7) &	 28.09 &	 0.11 &	 3 &	 0.66 &	 32.8 &	 0.36(7) &	 11.8 &	 19 \\ 
1843$-$0050 &	 18:43:36.72(6) &	 $-$00:50:10(3) &	 31.30 &	 1.36 &	 5 &	 0.99 &	 15.7 &	 0.24(5) &	 19.1 &	 -- \\ 
& & & & & & & & & & \\ 
1843$-$0355 &	 18:43:06.67(5) &	 $-$03:55:56(3) &	 28.48 &	 0.06 &	 2 &	 0.65 &	 16.4 &	 0.8(2) &	 16.7 &	 -- \\ 
1843$-$0459 &	 18:43:27.64(2) &	 $-$04:59:30(1) &	 27.58 &	 $-$0.51 &	 4 &	 1.09 &	 50.6 &	 1.7(3) &	 56.0 &	 76 \\ 
1844$-$0310 &	 18:44:45.48(5) &	 $-$03:10:37(2) &	 29.34 &	 0.04 &	 9 &	 1.26 &	 10.8 &	 0.7(1) &	 20.4 &	 -- \\ 
1847$-$0438 &	 18:47:37.93(1) &	 $-$04:38:15.3(9) &	 28.37 &	 $-$1.27 &	 5 &	 0.83 &	 53.8 &	 0.5(1) &	 11.3 &	 -- \\ 
1847$-$0605 &	 18:47:21.07(1) &	 $-$06:05:14.1(9) &	 27.05 &	 $-$1.87 &	 12 &	 1.24 &	 40.5 &	 0.8(2) &	 13.1 &	 38 \\ 
& & & & & & & & & & \\ 
1849$+$0127$^{\rm A}$ &	 18:49:44.2(2) &	 $+$01:27:23(4) &	 34.03 &	 1.04 &	 3 &	 0.77 &	 20.2 &	 0.46(9) &	 17.0 &	 -- \\ 
1849$-$0317 &	 18:49:57.85(3) &	 $-$03:17:31(3) &	 29.83 &	 $-$1.17 &	 12 &	 0.20 &	 61.8 &	 0.7(1) &	 23.2 &	 33 \\ 
1850$+$0026 &	 18:50:45.14(1) &	 $+$00:26:25.6(9) &	 33.25 &	 0.35 &	 6 &	 0.97 &	 50.0 &	 1.0(2) &	 14.1 &	 99 \\ 
1852$+$0305$^{\rm A}$ &	 18:52:32.6(5) &	 $+$03:05:05(9) &	 35.80 &	 1.16 &	 7 &	 1.26 &	 11.2 &	 0.8(2) &	 15.6 &	 -- \\ 
1853$+$0056$^{\rm A}$ &	 18:53:32.7(2) &	 $+$00:56:59(4) &	 34.02 &	 $-$0.04 &	 3 &	 0.38 &	 13.1 &	 0.21(4) &	 5.4 &	 -- \\ 
\hline
\end{tabular}
\end{minipage}
\end{table*}

%% file: mb1069tb1c.tex
\begin{table*}
\begin{minipage}{150mm}
\caption{-- {\it continued}}
\begin{tabular}{lllrrccrllr}
\hline 
\multicolumn{1}{c}{PSR J} & R.A. (J2000) & Dec. (J2000) & 
\multicolumn{1}{c}{$l$} & \multicolumn{1}{c}{$b$} & Beam & Radial & 
\multicolumn{1}{c}{$S/N$} & \multicolumn{1}{c}{$S_{1400}$} & \multicolumn{1}{c} {$W_{50}$} & 
\multicolumn{1}{c}{$W_{10}$} \\ & (h~~~m~~~s) & (~\degr ~~~\arcmin ~~~\arcsec) & 
\multicolumn{1}{c}{(\degr)} & \multicolumn{1}{c}{(\degr)} &   & Dist. &  & 
\multicolumn{1}{c}{(mJy)} & \multicolumn{1}{c}{(ms)} & \multicolumn{1}{c}{(ms)} \\ 
\hline 
1855$+$0422 &	 18:55:41.4(1) &	 $+$04:22:47(4) &	 37.31 &	 1.05 &	 1 &	 1.40 &	 13.8 &	 0.45(9) &	 42.7 &	 -- \\ 
1856$+$0404$^{\rm A}$ &	 18:56:26.6(3) &	 $+$04:04:26(5) &	 37.13 &	 0.75 &	 4 &	 0.48 &	 22.1 &	 0.48(10) &	 6.6 &	 -- \\ 
1857$+$0210 &	 18:57:40.90(4) &	 $+$02:10:58(3) &	 35.59 &	 $-$0.39 &	 7 &	 0.90 &	 15.1 &	 0.30(6) &	 16.0 &	 -- \\ 
1858$+$0215 &	 18:58:17.4(1) &	 $+$02:15:38(8) &	 35.72 &	 $-$0.49 &	 7 &	 0.64 &	 15.3 &	 0.22(4) &	 32.5 &	 -- \\ 
1900$+$0227 &	 19:00:38.60(4) &	 $+$02:27:32(2) &	 36.17 &	 $-$0.92 &	 5 &	 1.10 &	 20.0 &	 0.33(7) &	 17.3 &	 -- \\ 
& & & & & & & & & & \\ 
1901$+$0413$^{\rm A}$ &	 19:01:10.3(1) &	 $+$04:13:51(6) &	 37.81 &	 $-$0.23 &	 1 &	 1.11 &	 38.9 &	 1.1(2) &	 83.5 &	 -- \\ 
1904$+$0412$^{\rm P,~A}$ &	 19:04:31.382(4) &	 $+$04:12:05.9(1) &	 38.16 &	 $-$0.99 &	 10 &	 0.16 &	 13.4 &	 0.23(5) &	 2.5 &	 -- \\ 
1905$+$0616 &	 19:05:06.85(1) &	 $+$06:16:15.7(8) &	 40.07 &	 $-$0.17 &	 12 &	 0.20 &	 47.2 &	 0.40(8) &	 9.3 &	 20 \\ 
1906$+$0912 &	 19:06:28.46(3) &	 $+$09:12:57(1) &	 42.84 &	 0.88 &	 2 &	 1.02 &	 17.7 &	 0.32(6) &	 16.7 &	 -- \\ 
1907$+$0740 &	 19:07:44.12(1) &	 $+$07:40:22.6(9) &	 41.61 &	 $-$0.10 &	 7 &	 1.04 &	 29.5 &	 0.41(8) &	 13.2 &	 23 \\ 
& & & & & & & & & & \\ 
1907$+$0534$^{\rm A}$ &	 19:07:23.3(2) &	 $+$05:34:53(14) &	 39.72 &	 $-$0.99 &	 9 &	 0.87 &	 12.5 &	 0.36(7) &	 8.9 &	 -- \\ 
1908$+$0909 &	 19:08:07.44(1) &	 $+$09:09:12.4(5) &	 42.97 &	 0.49 &	 12 &	 0.56 &	 19.1 &	 0.22(4) &	 6.9 &	 -- \\ 
1908$+$0839$^{\rm A}$ &	 19:08:18.51(4) &	 $+$08:39:59(1) &	 42.56 &	 0.23 &	 7 &	 0.43 &	 26.8 &	 0.49(10) &	 5.1 &	 -- \\ 
1909$+$0912 &	 19:09:19.91(5) &	 $+$09:12:54(2) &	 43.16 &	 0.26 &	 5 &	 1.04 &	 17.1 &	 0.35(7) &	 10.0 &	 -- \\ 
1909$+$0616 &	 19:09:51.21(4) &	 $+$06:16:52(2) &	 40.62 &	 $-$1.21 &	 2 &	 0.32 &	 14.8 &	 0.33(7) &	 32.5 &	 -- \\ 
& & & & & & & & & & \\ 
1910$+$0534 &	 19:10:26.51(9) &	 $+$05:34:09(4) &	 40.06 &	 $-$1.67 &	 8 &	 0.45 &	 13.0 &	 0.41(8) &	 25.8 &	 -- \\ 
1913$+$1145 &	 19:13:43.88(6) &	 $+$11:45:33(1) &	 45.92 &	 0.48 &	 4 &	 0.69 &	 24.4 &	 0.43(9) &	 14.4 &	 -- \\ 
1913$+$1011 &	 19:13:20.341(3) &	 $+$10:11:22.97(7) &	 44.48 &	 $-$0.17 &	 9 &	 0.53 &	 19.0 &	 0.5(1) &	 2.1 &	 5 \\ 
1913$+$0832$^{\rm P,~A}$ &	 19:13:00.50(2) &	 $+$08:32:05.1(5) &
42.98 &	 $-$0.86 &	 11 &	 0.71 &	 23.7 &	 0.6(1) &	 77.0 &	 -- \\ 
1920$+$1110 &	 19:20:13.31(6) &	 $+$11:10:59(3) &	 46.15 &	 $-$1.20 &	 8 &	 0.73 &	 16.2 &	 0.39(8) &	 10.2 &	 -- \\ 
\hline
\end{tabular}
\end{minipage}
\end{table*}

%% file: mb1069tb2a.tex
\begin{table*}
\begin{minipage}{150mm}
\caption{Rotational parameters and dispersion measures for 120 pulsars
discovered in the Parkes multibeam pulsar survey.  Asterisks indicate
those pulsars whose
timing residuals contain significant timing noise which has been
removed, to first order, by the fitting of a period double derivative.}
\label{tb:per}
\begin{tabular}{lllcccrl}
\hline
\multicolumn{1}{c}{PSR J} & \multicolumn{1}{c}{Period, $P$} & \multicolumn{1}{c} {$\dot P$} &  
Epoch & $N_{\rm TOA}$ & Data Span & \multicolumn{1}{c}{Residual} & \multicolumn{1}{c}{DM}  \\ 
& \multicolumn{1}{c}{(s)} & \multicolumn{1}{c}{($10^{-15}$)} & 
(MJD) & & (MJD)  & \multicolumn{1}{c}{(ms)}  & \multicolumn{1}{c}{(cm$^{-3}$ pc)} \\ 
\hline 
0729$-$1448$^*$ & 0.251658714181(12)&	 113.2879(4) &	 51367.0 &	 112 &	50758$-$51975  &	 2.9 &	 92.3(3) \\ 
0737$-$2202$^*$ & 0.320366370453(3) &	 5.46883(9)  &	 51581.0 &	 163 &	50831$-$52199  &	 0.8 &	 95.7(4) \\ 
1733$-$3322 &	 1.24591463402(8) &	 4.10(3) &	 51250.0 &	 26 &	51026$-$51473  &	 1.6 &	 524.0(17) \\ 
1734$-$3333$^*$ & 1.1690082980(19) &	 2278.97(4) &	 51457.0 &	 34 &	51032$-$51881  &	 19.0 &	 578(9) \\ 
1735$-$3258 &	 0.35096322944(12) &	 26.08(7) &	 51573.0 &	 22 &	51393$-$51752  &	 4.1 &	 754(8) \\ 
& & & & & & & \\ 								              
1737$-$3102 &	 0.76867227339(5) &	 37.334(6) &	 51112.0 &	 31 &	50758$-$51464  &	 2.0 &	 280(6) \\ 
1737$-$3137$^*$ & 0.45043237000(4) &	 138.756(2) &	 51234.0 &	 53 &	50832$-$51517  &	 2.5 &	 488.2(10) \\ 
1738$-$2955$^*$ & 0.4433980218(3) &	 81.861(4) &	 51528.0 &	 20 &	51158$-$51898  &	 0.8 &	 223.4(6) \\ 
1739$-$3023$^*$ & 0.11436793182(2) &	 11.40145(12) &	 51497.0 &	 48 &	50727$-$52266  &	 1.5 &	 170.0(3) \\ 
1739$-$3049 &	 0.239317178406(18) &	 2.1781(18) &	 51116.0 &	 27 &	50758$-$51473  &	 2.0 &	 573(2) \\ 
& & & & & & & \\								              
1739$-$3159 &	 0.87756123714(15) &	 0.197(15) &	 51158.0 &	 20 &	50850$-$51464  &	 3.1 &	 337(5) \\ 
1740$-$3052 &	 0.570309580513(16) &	 2.54969(4) &	 51452.0 &	 139 &	51032$-$51872  &	 0.8 &	 740.9(2) \\ 
1741$-$2733 &	 0.89295866978(6) &	 0.148(8) &	 51184.0 &	 36 &	50904$-$51463  &	 1.6 &	 149.2(17) \\ 
1741$-$2945 &	 0.223557828541(15) &	 0.634(2) &	 51088.0 &	 31 &	50764$-$51410  &	 1.9 &	 310.3(12) \\ 
1741$-$3016 &	 1.89374869377(18) &	 8.99(6) &	 50959.0 &	 27 &	50728$-$51188  &	 2.3 &	 382(6) \\ 
& & & & & & & \\								              
1743$-$3153$^*$ &0.193105399932(6) &	 10.5674(3) &	 51309.0 &	 40 &	50879$-$51737  &	 0.9 &	 505.7(12) \\ 
1744$-$3130 &	 1.06606087128(4) &	 21.224(12) &	 50975.0 &	 26 &	50760$-$51188  &	 0.6 &	 192.9(7) \\ 
1747$-$2802 &	 2.7800791979(8) &	 2.37(12) &	 51312.0 &	 20 &	51037$-$51586  &	 6.2 &	 835(14) \\ 
1749$-$2629 &	 1.33538779057(13) &	 1.72(3) &	 51495.0 &	 16 &	51251$-$51738  &	 1.5 &	 409(11) \\ 
1750$-$2438 &	 0.712794036774(10) &	 10.797(3) &	 51491.0 &	 26 &	51242$-$51738  &	 0.3 &	 476(5) \\ 
& & & & & & & \\								              
1751$-$2516$^*$ & 0.39483578357(9) &	 2.644(2) &	 51383.0 &	 22 &	51013$-$51752  &	 1.8 &	 556(3) \\ 
1752$-$2821 &	 0.64022949952(4) &	 3.468(17) &	 50974.0 &	 20 &	50761$-$51185  &	 1.1 &	 516.3(13) \\ 
1755$-$2521$^*$ & 1.1759678772(3) &	 90.193(15) &	 51586.0 &	 37 &	51243$-$51985  &	 6.1 &	 252(4) \\ 
1755$-$2725 &	 0.26195467832(5) &	 0.014(9) &	 51492.0 &	 21 &	51243$-$51739  &	 3.4 &	 115(5) \\ 
1757$-$2223 &	 0.1853101015788(11) &	 0.7820(3) &	 51495.0 &	 26 &	51250$-$51739  &	 0.1 &	 239.3(4) \\ 
& & & & & & & \\								              
1758$-$2206 &	 0.430278250497(15) &	 0.956(4) &	 51495.0 &	 22 &	51250$-$51740  &	 0.8 &	 678(4) \\ 
1758$-$2540 &	 2.1072633712(4) &	 1.55(14) &	 50975.0 &	 22 &	50762$-$51187  &	 2.8 &	 218.2(13) \\ 
1758$-$2630 &	 1.20289346322(18) &	 5.17(8) &	 51244.0 &	 15 &	51033$-$51454  &	 2.2 &	 328(3) \\ 
1759$-$1940 &	 0.254720352269(7) &	 0.093(2) &	 51491.0 &	 25 &	51242$-$51740  &	 0.7 &	 302.7(10) \\ 
1759$-$1956 &	 2.8433888097(3) &	 18.57(5) &	 51495.0 &	 22 &	51250$-$51740  &	 1.5 &	 236.4(19) \\ 
& & & & & & & \\								              
1759$-$2302 &	 0.81071772906(12) &	 10.744(12) &	 51115.0 &	 23 &	50767$-$51462  &	 3.7 &	 889.0(1) \\ 
1759$-$2307 &	 0.55888867754(3) &	 3.762(3) &	 51110.0 &	 36 &	50757$-$51462  &	 1.1 &	 812.6(15) \\ 
1759$-$2549 & 0.95654854873(11) &	 99.592(13) &	 51380.0 &	 34 &	51021$-$51737  &	 4.9 &	 431(5) \\ 
1801$-$1855 &	 2.5504982104(10) &	 0.2(3) &	 51495.0 &	 15 &	51250$-$51740  &	 6.0 &	 484(14) \\ 
1801$-$1909 &	 1.10872484960(5) &	 0.703(9) &	 51491.0 &	 26 &	51242$-$51740  &	 1.0 &	 264(9) \\ 
& & & & & & & \\								              
1802$-$1745 &	 0.51467137652(3) &	 0.564(11) &	 51719.0 &	 24 &	51509$-$51929  &	 0.9 &	 264.2(3) \\ 
1802$-$2426 &	 0.56900709018(3) &	 8.562(7) &	 51495.0 &	 29 &	51250$-$51739  &	 1.3 &	 711(6) \\ 
1803$-$1857 &	 2.86433773648(13) &	 15.17(4) &	 51491.0 &	 24 &	51243$-$51739  &	 1.2 &	 392.0(11) \\ 
1804$-$2228 &	 0.57051063348(10) &	 0.143(10) &	 51414.0 &	 25 &	51087$-$51740  &	 3.5 &	 424(7) \\ 
1805$-$2032 &	 0.40576948790(5) &	 8.394(7) &	 51149.0 &	 19 &	50834$-$51462  &	 2.5 &	 932(2) \\ 
& & & & & & & \\								              
1805$-$2037 &	 0.357806774953(13) &	 1.7555(16) &	 51388.0 &	 30 &	51037$-$51739  &	 1.5 &	 708.1(16) \\ 
1806$-$1920$^*$ & 0.8797908757(6) &	 0.02(2) &	 51271.0 &	 28 &	50802$-$51739  &	 13.5 &	 683(7) \\ 
1806$-$2125$^*$ &  0.48178844602(5) &	 117.295(14) &	 51063.0 &	 37 &	50820$-$51305  &	 3.5 &	 750(3) \\ 
1809$-$1429 &	 0.895285284345(13) &	 5.240(4) &	 51495.0 &	 26 &	51250$-$51740  &	 0.4 &	 411.3(16) \\ 
1809$-$1917$^*$ & 0.082746885745(2) &	 25.53546(3) &	 51506.0 &	 50 &	50820$-$52191  &	 0.5 &	 197.1(4) \\ 
& & & & & & & \\								              
1809$-$2004 &	 0.43481138633(16) &	 7.28(6) &	 51719.0 &	 21 &	51509$-$51928  &	 5.3 &	 867.1(17) \\ 
1810$-$1820 &	 0.153716278369(8) &	 0.052(3) &	 50972.0 &	 35 &	50757$-$51186  &	 1.1 &	 452(3) \\ 
1810$-$2005 &	 0.03282224432571(9) &	 0.000151(7) &	 51200.0 &	 269 &	50757$-$51632  &	 0.4 &	 240.2(3) \\ 
1811$-$1736 &	 0.104181954734(3) &	 0.0018(6) &	 51050.0 &	 475 &	50801$-$51928  &	 1.0 &	 477(10) \\ 
1811$-$1835 &	 0.5574636120(1) &	 6.315(13) &	 51184.0 &	 22 &	50905$-$51463  &	 2.2 &	 761(11) \\ 
\hline
\end{tabular}
\end{minipage}
\end{table*}

%% file: mb1069tb2b.tex
\begin{table*}
\begin{minipage}{150mm}
\caption{-- {\it continued}}
\begin{tabular}{lllcccrl}
\hline
\multicolumn{1}{c}{PSR J} & \multicolumn{1}{c}{Period, $P$} & \multicolumn{1}{c} {$\dot P$} &  
Epoch & $N_{\rm TOA}$ & Data Span & \multicolumn{1}{c}{Residual} & \multicolumn{1}{c}{DM}  \\ 
& \multicolumn{1}{c}{(s)} & \multicolumn{1}{c}{($10^{-15}$)} & 
(MJD) & & (MJD)  & \multicolumn{1}{c}{(ms)}  & \multicolumn{1}{c}{(cm$^{-3}$ pc)} \\ 
\hline 
1812$-$2102 &	 1.22335230807(9) &	 23.893(9) &	 51134.0 &	 32 &	50802$-$51464  &	 1.5 &	 547.2(10) \\ 
1813$-$2113 &	 0.426466240565(13) &	 2.081(6) &	 51018.0 &	 29 &	50802$-$51233  &	 0.6 &	 462.3(15) \\ 
1814$-$1649 &	 0.95746393536(7) &	 6.333(7) &	 51133.0 &	 25 &	50801$-$51463  &	 1.5 &	 782(6) \\ 
1814$-$1744$^*$ & 3.9758447335(15) &	 743.05(4) &	 51413.0 &	 56 &	50850$-$51975  &	 14.8 &	 792(16) \\ 
1815$-$1910 &	 1.24992353738(8) &	 36.30(3) &	 50954.0 &	 23 &	50721$-$51186  &	 1.4 &	 547.8(4) \\ 
& & & & & & & \\								              
1818$-$1519 &	 0.9396900396(5) &	 4.11(7) &	 51415.0 &	 34 &	51091$-$51739  &	 15.0 &	 845(6) \\ 
1818$-$1541 &	 0.55113377674(3) &	 9.676(7) &	 51491.0 &	 25 &	51243$-$51739  &	 1.4 &	 690(5) \\ 
1819$-$1408 &	 1.7884904528(16) &	 2.59(16) &	 51612.0 &	 21 &	51295$-$51929  &	 20.0 &	 1075(41) \\ 
1819$-$1510 &	 0.226538990075(4) &	 0.0079(9) &	 51492.0 &	 25 &	51250$-$51733  &	 0.4 &	 421.7(11) \\ 
1823$-$1347 &	 0.61710720555(12) &	 9.60(3) &	 51481.0 &	 23 &	51250$-$51711  &	 3.1 &	 1044(6) \\ 
& & & & & & & \\								              
1823$-$1807 &	 1.63679245172(17) &	 0.28(4) &	 51295.0 &	 21 &	51036$-$51554  &	 2.0 &	 330(4) \\ 
1824$-$1159 &	 0.362492078224(8) &	 5.379(2) &	 51495.0 &	 28 &	51250$-$51739  &	 0.6 &	 463(2) \\ 
1824$-$1423 &	 0.359394162457(6) &	 0.3923(19) &	 51719.0 &	 27 &	51507$-$51929  &	 0.3 &	 428.3(11) \\ 
1826$-$1526 &	 0.382072812116(17) &	 1.085(8) &	 51720.0 &	 23 &	51509$-$51929  &	 1.0 &	 530(4) \\ 
1827$-$0934 &	 0.512547817784(16) &	 7.227(4) &	 51491.0 &	 27 &	51242$-$51740  &	 0.7 &	 259.2(5) \\ 
& & & & & & & \\								              
1828$-$1101$^*$ & 0.072051632709(2) &	 14.80952(9) &	 51531.0 &	 62 &	51088$-$51974  &	 0.9 &	 607.4(5) \\ 
1828$-$1336 &	 0.86033211599(7) &	 0.995(15) &	 51495.0 &	 24 &	51250$-$51740  &	 1.9 &	 494.7(18) \\ 
1830$-$1135$^*$ & 6.2215526664(12) &	 47.75(4) &	 51563.0 &	 34 &	51152$-$51974  &	 3.9 &	 257(6) \\ 
1831$-$1223 &	 2.8579410013(6) &	 5.5(3) &	 51704.0 &	 27 &	51477$-$51929  &	 4.5 &	 342(5) \\ 
1831$-$1329 &	 2.1656793503(3) &	 2.99(6) &	 51494.0 &	 21 &	51250$-$51737  &	 2.6 &	 338(5) \\ 
& & & & & & & \\								              
1832$-$0644 &	 0.74429541939(5) &	 37.089(17) &	 51489.0 &	 20 &	51250$-$51728  &	 1.4 &	 578(7) \\ 
1833$-$0559 &	 0.48345888008(5) &	 12.35(3) &	 51719.0 &	 21 &	51509$-$51928  &	 2.0 &	 353(6) \\ 
1833$-$1055 &	 0.63364029817(3) &	 0.527(4) &	 51360.0 &	 22 &	51088$-$51632  &	 1.1 &	 543(4) \\ 
1834$-$0602 &	 0.48791359170(3) &	 1.828(10) &	 51704.0 &	 25 &	51477$-$51929  &	 1.3 &	 445(4) \\ 
1835$-$0924 &	 0.8591920352(3) &	 21.39(9) &	 51720.0 &	 23 &	51509$-$51930  &	 4.7 &	 471(7) \\ 
& & & & & & & \\								              
1835$-$1020 &	 0.302448108644(5) &	 5.9187(11) &	 51493.0 &	 19 &	51250$-$51736  &	 0.3 &	 113.7(9) \\ 
1837$-$0559 &	 0.201062574029(5) &	 3.3048(13) &	 51463.0 &	 29 &	51186$-$51740  &	 0.7 &	 317.8(7) \\ 
1837$-$0604$^*$ &  0.09629420774(3) &	 45.1724(8) &	 51749.0 &	 73 &	51153$-$52344  &	 11.5 &	 462(1) \\ 
1837$-$1243 &	 1.87601852435(11) &	 36.51(3) &	 51494.0 &	 21 &	51250$-$51738  &	 1.1 &	 300(9) \\ 
1838$-$0453 &	 0.38083077683(3) &	 115.660(8) &	 51493.0 &	 24 &	51250$-$51735  &	 1.8 &	 621(9) \\ 
& & & & & & & \\								              
1838$-$1046 &	 1.21835359632(8) &	 3.080(18) &	 51493.0 &	 19 &	51250$-$51736  &	 1.2 &	 208(3) \\ 
1839$-$0321 &	 0.238777817900(9) &	 12.520(3) &	 51489.0 &	 24 &	51250$-$51728  &	 0.8 &	 449(2) \\ 
1839$-$0402 &	 0.520939679547(17) &	 7.694(5) &	 51487.0 &	 27 &	51250$-$51724  &	 0.9 &	 242(3) \\ 
1839$-$0459 &	 0.58531903782(3) &	 3.308(5) &	 51426.0 &	 23 &	51152$-$51699  &	 1.1 &	 243(3) \\ 
1839$-$06273 &	 0.48491367908(4) &	 0.132(17) &	 51361.0 &	 16 &	51153$-$51568  &	 1.0 &	 88.5(7) \\ 
& & & & & & & \\								              
1839$-$0643 &	 0.449548149848(15) &	 3.638(7) &	 51368.0 &	 20 &	51154$-$51581  &	 0.5 &	 497.9(16) \\ 
1841$-$0348 &	 0.204068115056(17) &	 57.872(3) &	 51646.0 &	 29 &	51359$-$51931  &	 1.9 &	 194.2(4) \\ 
1842$-$0153 &	 1.05422825038(3) &	 6.722(4) &	 51432.0 &	 27 &	51152$-$51710  &	 0.7 &	 434(5) \\ 
1842$-$0415 &	 0.526682241575(12) &	 21.9366(12) &	 51589.0 &	 31 &	51250$-$51928  &	 0.8 &	 194(6) \\ 
1843$-$0050 &	 0.78259846861(8) &	 0.249(12) &	 51462.0 &	 21 &	51184$-$51740  &	 2.2 &	 507(7) \\ 
& & & & & & & \\								              
1843$-$0355 &	 0.132313618371(11) &	 1.040(3) &	 51490.0 &	 28 &	51250$-$51729  &	 2.0 &	 798(3) \\ 
1843$-$0459 &	 0.75496337007(3) &	 0.854(10) &	 51719.0 &	 22 &	51507$-$51930  &	 0.7 &	 444.1(5) \\ 
1844$-$0310 &	 0.52504914274(4) &	 10.235(4) &	 51593.9 &	 53 &	51257$-$51931  &	 3.3 &	836(7)  \\ 
1847$-$0438 &	 0.95799116473(3) &	 10.933(9) &	 51319.0 &	 24 &	51092$-$51544  &	 0.5 &	 229(4) \\ 
1847$-$0605 &	 0.778164353310(19) &	 4.645(7) &	 51719.0 &	 23 &	51507$-$51930  &	 0.5 &	 207.9(18) \\ 
& & & & & & & \\								              
1849$+$0127 &	 0.54215548135(9) &	 27.97(6) &	 51664.0 &	 18 &	51478$-$51848  &	 2.1 &	 207(3) \\ 
1849$-$0317 &	 0.66840782878(4) &	 22.030(18) &	 51720.0 &	 21 &	51509$-$51931  &	 1.1 &	 42.9(28) \\ 
1850$+$0026 &	 1.08184378189(3) &	 0.359(11) &	 51353.0 &	 26 &	51140$-$51564  &	 0.5 &	 201.4(12) \\ 
1852$+$0305 &	 1.3261485670(8) &	 0.1(6) &	 51664.0 &	 18 &	51479$-$51848  &	 6.2 &	 320(12) \\ 
1853$+$0056 &	 0.27557759958(6) &	 21.39(4) &	 51665.0 &	 20 &	51480$-$51848  &	 2.5 &	 180.9(12) \\ 
\hline
\end{tabular}
\end{minipage}
\end{table*}

%% file: mb1069tb2c.tex
\begin{table*}
\begin{minipage}{150mm}
\caption{-- {\it continued}}
\begin{tabular}{lllcccrl}
\hline
\multicolumn{1}{c}{PSR J} & \multicolumn{1}{c}{Period, $P$} & \multicolumn{1}{c} {$\dot P$} &  
Epoch & $N_{\rm TOA}$ & Data Span & \multicolumn{1}{c}{Residual} & \multicolumn{1}{c}{DM}  \\ 
& \multicolumn{1}{c}{(s)} & \multicolumn{1}{c}{($10^{-15}$)} & 
(MJD) & & (MJD)  & \multicolumn{1}{c}{(ms)}  & \multicolumn{1}{c}{(cm$^{-3}$ pc)} \\ 
\hline 
1855$+$0422 &	 1.6781063295(3) &	 0.93(9) &	 51446.0 &	 22 &	51184$-$51707  &	 4.1 &	 438(6) \\ 
1856$+$0404 &	 0.42025215518(9) &	 0.04(6) &	 51664.0 &	 25 &	51479$-$51848  &	 3.7 &	 341.3(7) \\ 
1857$+$0210 &	 0.63098305833(6) &	 14.03(2) &	 51721.0 &	 22 &	51509$-$51931  &	 1.6 &	 783(11) \\ 
1858$+$0215 &	 0.7458280320(3) &	 4.61(9) &	 51720.0 &	 19 &	51509$-$51929  &	 5.3 &	 702.0(1) \\ 
1900$+$0227 &	 0.37426157516(3) &	 5.705(12) &	 51721.0 &	 25 &	51509$-$51931  &	 1.5 &	 201.1(17) \\ 
& & & & & & & \\								              
1901$+$0413 &	 2.6630796830(8) &	 131.6(3) &	 51706.0 &	 37 &	51479$-$51931  &	 5.4 &	 352(3) \\ 
1904$+$0412 &	 0.0710948973807(3) &	 0.00011(3) &	 51450.0 &	 65 &	51089$-$51804  &	 0.2 &	 185.9(7) \\ 
1905$+$0616 &	 0.98970706304(3) &	 135.218(11) &	 51720.0 &	 26 &	51507$-$51931  &	 0.7 &	 259(7) \\ 
1906$+$0912 &	 0.77534458995(5) &	 0.132(19) &	 51360.0 &	 19 &	51140$-$51579  &	 1.0 &	 265(5) \\ 
1907$+$0740 &	 0.574697951954(17) &	 0.671(7) &	 51354.0 &	 30 &	51138$-$51568  &	 0.7 &	 332(3) \\ 
& & & & & & & \\								              
1907$+$0534 &	 1.1384027132(6) &	 3.15(12) &	 51719.0 &	 24 &	51480$-$51957  &	 10.5 &	 524(4) \\ 
1908$+$0909$^*$ & 0.336554651762(14) &	 34.8712(6) &	 51525.0 &	 43 &	51138$-$51911  &	 0.9 &	 467.5(15) \\ 
1908$+$0839 &	 0.185397243919(14) &	 2.3865(14) &	 51550.0 &	 31 &	51251$-$51848  &	 2.1 &	 512(2) \\ 
1909$+$0912 &	 0.222949273318(15) &	 35.805(5) &	 51496.0 &	 27 &	51252$-$51740  &	 1.8 &	 421.5(17) \\ 
1909$+$0616 &	 0.75599276085(18) &	 20.58(8) &	 51720.0 &	 27 &	51508$-$51930  &	 1.7 &	 352(4) \\ 
& & & & & & & \\								              
1910$+$0534 &	 0.45286735393(9) &	 1.92(4) &	 51709.0 &	 24 &	51507$-$51909  &	 3.2 &	 484(3) \\ 
1913$+$1145 &	 0.30606864497(3) &	 5.016(11) &	 51290.0 &	 27 &	51094$-$51484  &	 1.4 &	 637(2) \\ 
1913$+$1011 &	 0.03590861279827(15) &	 3.36768(6) &	 51697.0 &	 88 &	51464$-$51929  &	 0.2 &	 178.8(3) \\ 
1913$+$0832$^*$ & 0.134409003573(10) &	 4.5696(6) &	 51685.0 &	 54 &	51223$-$52147  &	 1.1 &	 355.2(10) \\ 
1920$+$1110 &	 0.50988582045(5) &	 0.156(19) &	 51719.0 &	 15 &	51507$-$51929  &	 1.6 &	 182(3) \\ 
\hline
\end{tabular}
\end{minipage}
\end{table*}

%% file: mb1069tb4a.tex
\begin{table*}
\begin{minipage}{150mm}
\caption{Derived parameters for 120 newly discovered pulsars.  For
pulsars in which the measured $\dot{P}$ is not significant, only
limits are given for $\tau_c$, $B_s$ and $\dot E$.}
\label{tb:deriv}
\begin{tabular}{lccccrc}
\hline
\multicolumn{1}{c}{PSR J} & log[$\tau_c$ (yr)] & log[$B_s$ (G)] & log[$\dot E$ (erg s$^{-1}$)] 
& $d$ & $z$!! & Luminosity \\ 
& & & & (kpc) & (kpc) & (mJy kpc$^2$) \\ 
\hline 
0729$-$1448 &	 4.55 &	 12.73 &	 35.45 &	 !4.3 &	 0.11    & !12.0 \\ 
0737$-$2202 &	 5.97 &	 12.13 &	 33.82 &	 !4.2 &	 $-$0.02 & !!8.3 \\ 
1733$-$3322 &	 6.68 &	 12.36 &	 31.92 &	 !6.3 &	 $-$0.03 & !31.4 \\ 
1734$-$3333 &	 3.91 &	 13.72 &	 34.75 &	 !7.5 &	 $-$0.06 & !29.3 \\ 
1735$-$3258 &	 5.33 &	 12.49 &	 34.38 &	 11.1 &	 $-$0.07 & !56.7 \\ 
& & & & & &  \\ 
1737$-$3102 &	 5.51 &	 12.73 &	 33.51 &	 !4.3 &	 0.03 &	 !11.5 \\ 
1737$-$3137 &	 4.71 &	 12.90 &	 34.78 &	 !5.8 &	 0.01 &	 !26.2 \\ 
1738$-$2955 &	 4.93 &	 12.79 &	 34.57 &	 !3.9 &	 0.05 &	 !!4.4 \\ 
1739$-$3023 &	 5.20 &	 12.06 &	 35.48 &	 !3.4 &	 0.02 &	 !11.7 \\ 
1739$-$3049 &	 6.24 &	 11.86 &	 33.80 &	 !7.0 &	 0.02 &	 !26.5 \\ 
& & & & & &  \\ 
1739$-$3159 &	 7.85 &	 11.62 &	 31.08 &	 !5.0 &	 $-$0.05 & !24.0 \\ 
1740$-$3052 &	 5.55 &	 12.59 &	 33.73 &	 10.8 &	 $-$0.03 & !81.6 \\ 
1741$-$2733 &	 7.98 &	 11.56 &	 30.91 &	 !3.3 &	 0.09 &	 !12.0 \\ 
1741$-$2945 &	 6.75 &	 11.58 &	 33.34 &	 !4.7 &	 0.03 &	 !12.8 \\ 
1741$-$3016 &	 6.52 &	 12.62 &	 31.72 &	 !5.1 &	 0.01 &	 !58.5 \\ 
& & & & & &  \\ 
1743$-$3153 &	 5.46 &	 12.16 &	 34.76 &	 !8.0 &	 $-$0.15 & !32.0 \\ 
1744$-$3130 &	 5.90 &	 12.68 &	 32.84 &	 !3.7 &	 $-$0.07 & !!9.7 \\ 
1747$-$2802 &	 7.27 &	 12.41 &	 30.63 &	 11.4 &	 0.02 &	 !65.0 \\ 
1749$-$2629 &	 7.09 &	 12.18 &	 31.45 &	 !5.4 &	 0.05 &	 !21.3 \\ 
1750$-$2438 &	 6.02 &	 12.45 &	 33.08 &	 !7.2 &	 0.15 &	 !27.0 \\ 
& & & & & &  \\ 
1751$-$2516 &	 6.37 &	 12.01 &	 33.23 &	 !7.5 &	 0.09 &	 !12.4 \\ 
1752$-$2821 &	 6.47 &	 12.18 &	 32.72 &	 !7.5 &	 $-$0.13 & !18.0 \\ 
1755$-$2521 &	 5.31 &	 13.02 &	 33.34 &	 !4.2 &	 $-$0.01 & !13.1 \\ 
1755$-$2725 &	 $>$8.48 & $<$10.78 &	$<$31.48 &	 !2.8 &	 $-$0.06 & !!4.3 \\ 
1757$-$2223 &	 6.57 &	 11.59 &	 33.69 &	 !4.1 &	 0.07 &	 !19.0 \\ 
& & & & & &  \\ 
1758$-$2206 &	 6.85 &	 11.81 &	 32.67 &	 11.5 &	 0.19 &	 !54.2 \\ 
1758$-$2540 &	 7.33 &	 12.26 &	 30.82 &	 !3.7 &	 $-$0.05 & !!8.9 \\ 
1758$-$2630 &	 6.57 &	 12.40 &	 32.08 &	 !5.0 &	 $-$0.11 & !10.2 \\ 
1759$-$1940 &	 7.64 &	 11.19 &	 32.34 &	 !5.2 &	 0.17 &	 !26.0 \\ 
1759$-$1956 &	 6.38 &	 12.87 &	 31.51 &	 !4.0 &	 0.14 &	 !!6.6 \\ 
& & & & & &  \\ 
1759$-$2302 &	 6.08 &	 12.48 &	 32.90 &	 11.8 &	 0.05 &	 185.2 \\ 
1759$-$2307 &	 6.37 &	 12.17 &	 32.93 &	 11.2 &	 0.05 &	 !84.0 \\ 
1759$-$2549 &	 5.18 &	 12.99 &	 33.65 &	 !5.9 &	 $-$0.11 & !20.9 \\ 
1801$-$1855 &	$>$7.91 & $<$12.06 &	 $<$30.88 &	 11.2 &	 0.41 &	 !59.0 \\ 
1801$-$1909 &	 7.40 &	 11.95 &	 31.30 &	 !4.9 &	 0.15 &	 !12.5 \\ 
& & & & & &  \\ 
1802$-$1745 &	 7.16 &	 11.74 &	 32.20 &	 !5.2 &	 0.22 &	 !!5.7 \\ 
1802$-$2426 &	 6.02 &	 12.35 &	 33.26 &	 12.2 &	 $-$0.19 & !90.8 \\ 
1803$-$1857 &	 6.48 &	 12.82 &	 31.40 &	 !6.2 &	 0.15 &	 !15.4 \\ 
1804$-$2228 &	 7.80 &	 11.46 &	 31.48 &	 !5.3 &	 $-$0.04 & !!5.6 \\ 
1805$-$2032 &	 5.88 &	 12.27 &	 33.70 &	 11.8 &	 0.06 &	 !98.9 \\ 
& & & & & &  \\ 
1805$-$2037 &	 6.51 &	 11.90 &	 33.18 &	 !9.3 &	 0.05 &	 !29.4 \\ 
1806$-$1920 &	$>$8.45 &$<$11.33 &	 $<$30.46 &	 10.4 &	 0.14 &	 204.4 \\ 
1806$-$2125 &	 4.81 &	 12.88 &	 34.61 &	 !9.9 &	 $-$0.04 & 105.9 \\ 
1809$-$1429 &	 6.43 &	 12.34 &	 32.46 &	 11.7 &	 0.50 &	 !86.2 \\ 
1809$-$1917 &	 4.71 &	 12.17 &	 36.26 &	 !3.7 &	 0.01 &	 !34.6 \\ 
& & & & & &  \\ 
1809$-$2004 &	 5.98 &	 12.26 &	 33.54 &	 10.9 &	 $-$0.04 & 104.6 \\ 
1810$-$1820 &	 7.67 &	 10.96 &	 32.76 &	 !5.6 &	 0.03 &	 !22.9 \\ 
1810$-$2005 &	 9.60 &	 9.32  &	 32.18 &	 !4.0 &	 $-$0.04 &!31.4 \\ 
1811$-$1736 &	 9.21 &	 10.02 &	 31.56 &	 !5.9 &	 0.05 &	 !44.9 \\ 
1811$-$1835 &	 6.15 &	 12.28 &	 33.15 &	 !9.5 &	 0.01 &	 !37.9 \\ 
\hline
\end{tabular}
\end{minipage}
\end{table*}

%% file: mb1069tb4b.tex
\begin{table*}
\begin{minipage}{150mm}
\caption{-- {\it continued}}
\begin{tabular}{lccccrc}
\hline
\multicolumn{1}{c}{PSR J} & log[$\tau_c$ (yr)] & log[$B_s$ (G)] & log[$\dot E$ (erg s$^{-1}$)] 
& $d$ & $z$!! & Luminosity \\ 
& & & & (kpc) & (kpc) & (mJy kpc$^2$) \\ 
\hline 
1812$-$2102 &	 5.91 &	 12.74 &	 32.72 &	 !9.5 &	 $-$0.22 & 130.0 \\ 
1813$-$2113 &	 6.51 &	 11.98 &	 33.04 &	 !8.7 &	 $-$0.25 & !47.7 \\ 
1814$-$1649 &	 6.38 &	 12.40 &	 32.45 &	 !9.6 &	 0.04 &	 !101.4 \\ 
1814$-$1744 &	 4.93 &	 13.74 &	 32.67 &	 10.2 &	 $-$0.04 & !73.9 \\ 
1815$-$1910 &	 5.74 &	 12.83 &	 32.86 &	 !8.3 &	 $-$0.13 & !22.0 \\ 
& & & & & &  \\ 
1818$-$1519 &	 6.56 &	 12.30 &	 32.30 &	 10.0 &	 0.03 &	 208.0 \\ 
1818$-$1541 &	 5.96 &	 12.37 &	 33.36 &	 !8.3 &	 $-$0.01 & !70.3 \\ 
1819$-$1408 &	 7.04 &	 12.34 &	 31.26 &	 12.5 &	 0.09 &	 !79.7 \\ 
1819$-$1510 &	 8.66 &	 10.63 &	 31.43 &	 !5.6 &	 $-$0.01 & !19.4 \\ 
1823$-$1347 &	 6.01 &	 12.39 &	 33.20 &	 11.1 &	 $-$0.04 & !50.5 \\ 
& & & & & &  \\ 
1823$-$1807 &	 7.97 &	 11.83 &	 30.40 &	 !6.4 &	 $-$0.25 & !16.0 \\ 
1824$-$1159 &	 6.03 &	 12.15 &	 33.65 &	 !6.1 &	 0.03 &	 !26.8 \\ 
1824$-$1423 &	 7.16 &	 11.58 &	 32.52 &	 !6.1 &	 $-$0.08 & !31.3 \\ 
1826$-$1526 &	 6.75 &	 11.81 &	 32.89 &	 10.9 &	 $-$0.29 & !54.7 \\ 
1827$-$0934 &	 6.05 &	 12.29 &	 33.32 &	 !4.5 &	 0.07 &	 !5.9 \\ 
& & & & & &  \\ 
1828$-$1101 &	 4.89 &	 12.02 &	 36.20 &	 !7.2 &	 0.01 &	 149.8 \\ 
1828$-$1336 &	 7.14 &	 11.97 &	 31.79 &	 !7.8 &	 $-$0.17 & !37.7 \\ 
1830$-$1135 &	 6.31 &	 13.24 &	 30.89 &	 !4.5 &	 $-$0.04 & !22.3 \\ 
1831$-$1223 &	 6.92 &	 12.60 &	 30.97 &	 !5.3 &	 $-$0.12 & !34.3 \\ 
1831$-$1329 &	 7.06 &	 12.41 &	 31.08 &	 !6.4 &	 $-$0.21 & !20.9 \\ 
& & & & & &  \\ 
1832$-$0644 &	 5.50 &	 12.73 &	 33.56 &	 !9.0 &	 0.16 &	 !52.7 \\ 
1833$-$0559 &	 5.79 &	 12.39 &	 33.63 &	 !5.8 &	 0.13 &	 !18.5 \\ 
1833$-$1055 &	 7.28 &	 11.77 &	 31.91 &	 !8.4 &	 $-$0.17 & !36.0 \\ 
1834$-$0602 &	 6.63 &	 11.98 &	 32.79 &	 !6.5 &	 0.11 &	 !33.4 \\ 
1835$-$0924 &	 5.80 &	 12.64 &	 33.11 &	 !6.3 &	 $-$0.09 & !20.6 \\ 
& & & & & &  \\ 
1835$-$1020 &	 5.91 &	 12.13 &	 33.92 &	 !2.6 &	 $-$0.06 & !13.0 \\ 
1837$-$0559 &	 5.98 &	 11.92 &	 34.20 &	 !5.0 &	 0.03 &	 !11.5 \\ 
1837$-$0604 &	 4.53 &	 12.32 &	 36.30 &	 !6.2 &	 0.03 &	 !28.1 \\ 
1837$-$1243 &	 5.91 &	 12.92 &	 32.34 &	 !8.1 &	 $-$0.35 & !11.2 \\ 
1838$-$0453 &	 4.72 &	 12.83 &	 34.92 &	 !8.2 &	 0.10 &	 !22.2 \\ 
& & & & & &  \\ 
1838$-$1046 &	 6.80 &	 12.29 &	 31.83 &	 !4.3 &	 $-$0.16 & !!9.2 \\ 
1839$-$0321 &	 5.48 &	 12.24 &	 34.56 &	 !6.9 &	 0.13 &	 !12.9 \\ 
1839$-$0402 &	 6.03 &	 12.31 &	 33.32 &	 !4.6 &	 0.06 &	 !!4.4 \\ 
1839$-$0459 &	 6.45 &	 12.15 &	 32.81 &	 !4.5 &	 0.03 &	 !!6.1 \\ 
1839$-$06273&    7.77 &	 11.41 &	 31.66 &	 !2.4 &$-$0.01& !!1.7 \\ 
& & & & & &  \\ 
1839$-$0643 &	 6.29 &	 12.11 &	 33.20 &	 !6.5 &	 $-$0.04 & !57.0 \\ 
1841$-$0348 &	 4.75 &	 12.54 &	 35.43 &	 !4.2 &	 0.03 &	 !25.4 \\ 
1842$-$0153 &	 6.40 &	 12.43 &	 32.36 &	 !6.3 &	 0.12 &	 !22.2 \\ 
1842$-$0415 &	 5.58 &	 12.54 &	 33.77 &	 !4.0 &	 0.01 &	 !!5.8 \\ 
1843$-$0050 &	 7.70 &	 11.65 &	 31.32 &	 !7.8 &	 0.18 &	 !14.6 \\ 
& & & & & &  \\ 
1843$-$0355 &	 6.30 &	 11.57 &	 34.26 &	 !8.8 &	 0.01 &	 !62.0 \\ 
1843$-$0459 &	 7.15 &	 11.91 &	 31.89 &	 !6.3 &	 $-$0.05 & !67.9 \\ 
1844$-$0310 &	 5.91 &	 12.37 &	 33.45 &	 !8.9 &	 0.01 &	 !52.3 \\ 
1847$-$0438 &	 6.14 &	 12.51 &	 32.69 &	 !4.8 &	 $-$0.10 & !11.1 \\ 
1847$-$0605 &	 6.42 &	 12.28 &	 32.59 &	 !4.5 &	 $-$0.15 & !15.8 \\ 
& & & & & &  \\ 
1849$+$0127 &	 5.49 &	 12.60 &	 33.84 &	 !4.7 &	 0.08 &	 !10.2 \\ 
1849$-$0317 &	 5.68 &	 12.59 &	 33.46 &	 !1.8 &	 $-$0.04 & !!2.1 \\ 
1850$+$0026 &	 7.68 &	 11.80 &	 31.04 &	 !4.2 &	 0.03 &	 !18.2 \\ 
1852$+$0305 &	 $>$7.48 & $<$ 11.99 &	 $<$31.07 &	 !6.7 &	 0.14 &	 !35.9 \\ 
1853$+$0056 &	 5.31 &	 12.39 &	 34.60 &	 !3.9 &	 $-$0.00 & !!3.2 \\ 
\hline
\end{tabular}
\end{minipage}
\end{table*}

%% file: mb1069tb4c.tex
\begin{table*}
\begin{minipage}{150mm}
\caption{-- {\it continued}}
\begin{tabular}{lccccrc}
\hline
\multicolumn{1}{c}{PSR J} & log[$\tau_c$ (yr)] & log[$B_s$ (G)] & log[$\dot E$ (erg s$^{-1}$)] 
& $d$ & $z$!! & Luminosity \\ 
& & & & (kpc) & (kpc) & (mJy kpc$^2$) \\ 
\hline 
1855$+$0422 &	 7.46 &	 12.10 &	 30.89 &	 !9.9 &	 0.18 &	 !44.1 \\ 
1856$+$0404 &	$>$7.82 & $<$11.32 &	$<$31.73&	 !7.0 &	 0.09 &	 !23.5 \\ 
1857$+$0210 &	 5.85 &	 12.48 &	 33.34 &	 15.4 &	 $-$0.10 & !71.1 \\ 
1858$+$0215 &	 6.41 &	 12.27 &	 32.64 &	 12.4 &	 $-$0.11 & !33.8 \\ 
1900$+$0227 &	 6.02 &	 12.17 &	 33.63 &	 !4.2 &	 $-$0.07 & !!5.8 \\ 
& & & & & &  \\ 
1901$+$0413 &	 5.51 &	 13.28 &	 32.45 &	 !7.1 &	 $-$0.03 & !54.9 \\ 
1904$+$0412 &	 9.43 &	 10.05 &	 31.04 &	 !4.0 &	 $-$0.07 & !!3.7 \\ 
1905$+$0616 &	 5.06 &	 13.07 &	 33.74 &	 !5.4 &	 $-$0.02 & !11.7 \\ 
1906$+$0912 &	 7.97 &	 11.51 &	 31.04 &	 !5.5 &	 0.09 &	 !!9.7 \\ 
1907$+$0740 &	 7.13 &	 11.80 &	 32.15 &	 !6.8 &	 $-$0.01 & !19.0 \\ 
& & & & & &  \\ 
1907$+$0534 &	 6.76 &	 12.28 &	 31.92 &	 12.0 &	 $-$0.21 & !51.8 \\ 
1908$+$0909 &	 5.18 &	 12.54 &	 34.56 &	 !8.8 &	 0.08 &	 !17.0 \\ 
1908$+$0839 &	 6.09 &	 11.83 &	 34.18 &	 !9.6 &	 0.04 &	 !45.2 \\ 
1909$+$0912 &	 4.99 &	 12.46 &	 35.11 &	 !8.2 &	 0.04 &	 !23.5 \\ 
1909$+$0616 &	 5.76 &	 12.60 &	 33.28 &	 !8.5 &	 $-$0.18 & !23.8 \\ 
& & & & & &  \\ 
1910$+$0534 &	 6.57 &	 11.98 &	 32.91 &	 15.5 &	 $-$0.45 & !98.5 \\ 
1913$+$1145 &	 5.99 &	 12.10 &	 33.84 &	 14.7 &	 0.12 &	 !92.9 \\ 
1913$+$1011 &	 5.23 &	 11.55 &	 36.46 &	 !4.5 &	 $-$0.01 & !10.1 \\ 
1913$+$0832 &	 5.67 &	 11.90 &	 34.87 &	 !7.8 &	 $-$0.12 & !37.7 \\ 
1920$+$1110 &	 7.71 &	 11.45 &	 31.66 &	 !5.0 &	 $-$0.10 & !!9.8 \\ 
\hline
\end{tabular}
\end{minipage}
\end{table*}